\documentclass[prl,twocolumn,preprintnumbers,superscriptaddress]{revtex4}
\usepackage{amsmath,amssymb,mathrsfs}
\usepackage[dvips]{graphicx}
\usepackage{epstopdf}
\usepackage{hyperref}
\usepackage{paralist}

\usepackage{subfigure}
\usepackage{dcolumn}
\usepackage{color}

\def\be{\begin{equation}}
\def\ee{\end{equation}}
\def\bea{\begin{eqnarray}}
\def\eea{\end{eqnarray}}
\def\nn{\nonumber}

\def\tbf{\textbf}

\def\ncsection#1{{\par{\vskip 5pt}\noindent\Large #1}}

\begin{document}

\def\papertitle{Formation and detection of a chiral orbital Bose liquid in an optical lattice}
\title{\papertitle}

\author{Xiaopeng Li}
\affiliation{Department of Physics and Astronomy, University of Pittsburgh,
  Pittsburgh, Pennsylvania 15260, USA} 
\affiliation{Kavli Institute for Theoretical Physics, University of California,
  Santa Barbara, CA 93106, USA} 
\author{Arun Paramekanti}
\affiliation{Department of Physics, University of Toronto, Toronto, Ontario M5S 1A7, Canada }
\affiliation{Canadian Institute for Advanced Research, Toronto, Ontario M5G 1Z8, Canada}
\author{Andreas Hemmerich}
\affiliation{Institut f\"ur Laser-Physik, Universit\"at Hamburg, Luruper Caussee 149, 22761 Hamburg, Germany} 
\author{W. Vincent Liu}
\affiliation{Department of Physics and Astronomy, University of Pittsburgh,
  Pittsburgh, Pennsylvania 15260, USA} 
\affiliation{Center for Cold Atom Physics, Chinese Academy of Sciences, Wuhan
  430071, China} 

\maketitle 

{\bf Recent experiments~\cite{2011_Wirth_pband, 2013_Oelschlaeger_pband} on
  $p$-orbital atomic bosons have suggested the emergence of a spectacular ultracold 
  superfluid with staggered orbital currents in optical lattices.  This raises
  fundamental questions like the effects of collective thermal fluctuations, and
  how to directly observe such chiral order.  Here, we show via Monte Carlo
  simulations that thermal fluctuations destroy this superfluid in an unexpected
  two-step process, unveiling an intermediate normal phase with
  spontaneously broken time-reversal symmetry, dubbed ``chiral Bose liquid''.
  For integer fillings ($n\geq 2$) in the chiral Mott regime~\cite{2011_Li_EFA},
  thermal fluctuations are captured by an effective orbital Ising model, and
  Onsager's powerful exact solution~\cite{1944_Onsager_IsingModel} is adopted to
  determine the transition from this intermediate liquid to the para-orbital normal
  phase at high temperature.  A suitable lattice quench is designed to convert
  the staggered angular momentum, previously thought by experts difficult to
  directly probe, into coherent orbital oscillations, providing a smoking-gun
  signature of chiral order.}

Orbital degrees of freedom and interactions play a crucial role in the emergence of many
complex phases in solid state materials. 
High temperature superconductivity in the
cuprates \cite{1986_HighTc} and pnictides \cite{2006_Pnictides_Kamihara}, 
colossal magnetoresistance observed in Mn oxides \cite{1993_CMR_vonHelmolt}, and
chiral p-wave superconductivity proposed in Sr$_2$RuO$_4$
\cite{1998_Sr214_Maeno}, are all nucleated by strong correlation 
effects in a multi-orbital setting~\cite{2000_Tokura_Nagaosa_orbital_Science}. 
For ultracold atomic gases, interaction effects combined with the band topology of 
$p$-orbitals~\cite{2011_Lewenstein_Liu_orbital_dance} has been argued to lead to exotic topological or superfluid (SF) 
phases for fermions~\cite{2008_Zhao_pmott,2010_Zhang_sppair,
2010_Cai_FFLO,2009_Hung_fpair,2011_Zhang_pdw,2011_Sun_TSM,2012_Li_orbitalladder_NatComm} as well as
bosons~\cite{2005_Isacsson_pband,2006_Liu_TSOC,2006_Kuklov_sf,2008_Lim_TSOC,2008_Vladimir_icsf,2010_Zhou_interband,
2011_Sengstock_honeycomb_BEC, 2011_Wirth_pband, 2011_Li_EFA,2011_Cai_TSOC,2012_Li_1Dpboson_PRL,
2013_Hebert_QMCpboson_PRB}.
Interactions are predicted to drive a semi-metal to topological 
insulator quantum phase transition in two dimensions (2D) for fermions in $p_x$, $p_y$ and $d_{x^2 -y^2}$ 
orbitals~\cite{2011_Sun_TSM}, while interacting $p$-orbital atomic fermions in 3D could lead to
axial orbital order~\cite{2011_Hauke_OrbFermion_PRA}. 
For weakly interacting 2D lattice bosons in $p_x$ and $p_y$-orbitals the ground state is proposed to be a 
SF with staggered $p_x\pm i p_y$ order~\cite{2006_Liu_TSOC};
such order is also found for 1D strongly interacting $p$-orbital bosons~\cite{2012_Li_1Dpboson_PRL}.
For bosons, these exotic phases can result from a particularly simple effect: repulsive contact interactions favor a maximization of the 
local angular momentum ${\cal L}_z$, a bosonic variant of the atomic Hund's rule for electrons~\cite{2006_Liu_TSOC, 2011_Li_EFA, 2013_Hebert_QMCpboson_PRB}.

While previous work has focused on the ground state properties of such
unconventional Bose SFs, here we address {two important outstanding issues. (i) How do thermal or quantum fluctuations, which are important in any experimental setting, melt these unconventional SF states? (ii) How can one directly detect the spatially modulated angular momentum underlying these unusual quantum states?}

\begin{figure}[htp]
\includegraphics[angle=0,width=\linewidth]{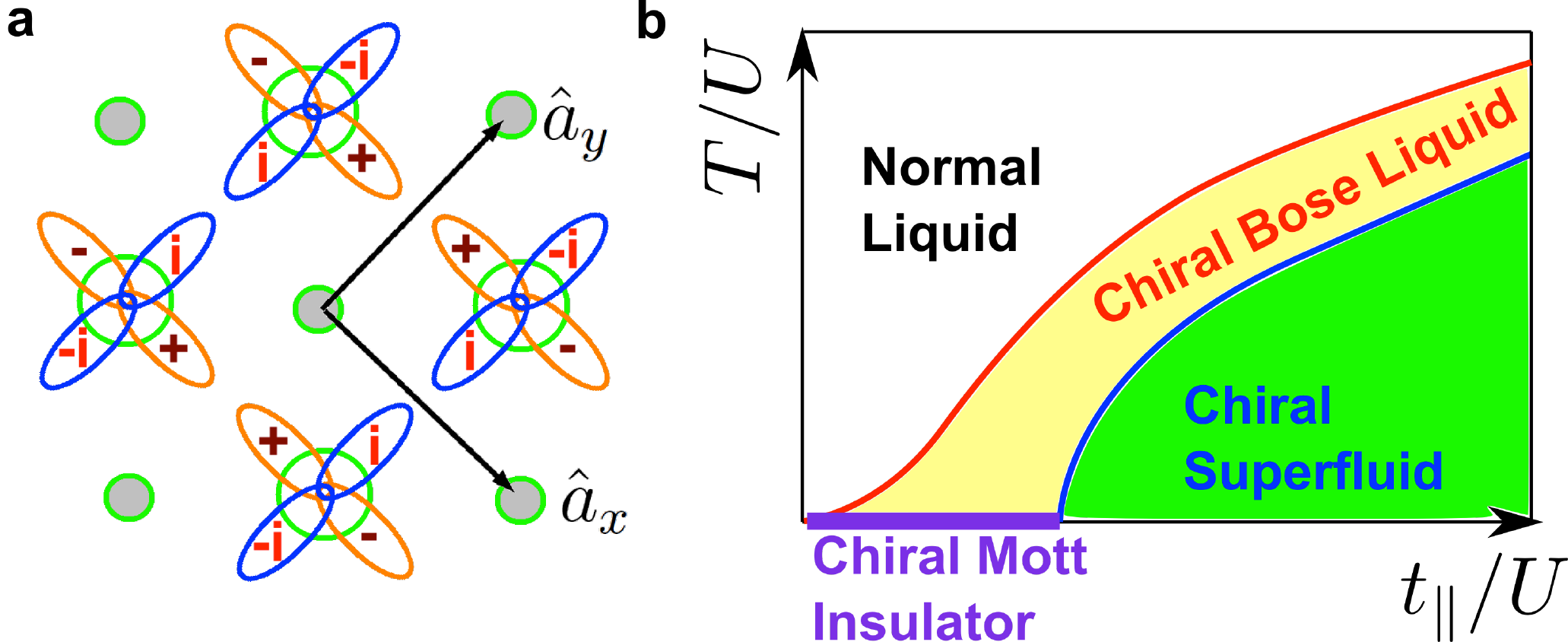}
\caption{{Orbital lattice structure and phase diagram. }
(\textbf{a}), the checkerboard lattice structure as used in experiments 
to realize $p$-orbital superfluid~\cite{2011_Wirth_pband}, together with
the phases of the staggered orbital ordering in chiral states.
(\textbf{b}) shows {the phase diagram of $p$-orbital band Bosons with filling 
$n\ge2$. The diagram is fixed by  exact or controlled  numerical
 and analytical calculations in the strong and weak coupling limits at non-zero
  temperature and for arbitrary coupling along the $T=0$ line of
  ``Chiral Mott'' and is otherwise  interpolated schematically elsewhere. }
At zero temperature there is a quantum phase transition between the chiral Mott
{insulator}
and superfluid phases, both of which break time reversal symmetry. 
At finite temperature, there is a chiral Bose liquid phase. Upon heating, the chiral superfluid 
undergoes a BKT transition into the chiral Bose liquid, which subsequently 
undergoes an Ising transition at a higher temperature into a normal Bose
liquid. 
}
\label{fig:latticephasediag}
\end{figure}

Our work is motivated by recent experiments which have successfully prepared 
long-lived metastable phases of weakly interacting $^{87}$Rb atoms in 
$p$-orbitals~\cite{2011_Wirth_pband, 2013_Oelschlaeger_pband}. In the 
deep lattice regime, this experimental system is well approximated by 
a tight binding model on a checkerboard optical lattice with bosons 
in the $p_x$, $p_y$ and $s$ orbital degrees of freedom (see Fig.~\ref{fig:latticephasediag}).  
The Hamiltonian of the model is obtained by extending the early theoretical 
studies~\cite{2005_Isacsson_pband, 2006_Liu_TSOC, 2009_Wu_pband, 2011_Li_EFA} to the checkerboard lattice configuration used 
in the recent experiments of Ref.~\cite{2011_Wirth_pband, 2013_Oelschlaeger_pband}. 
Restricting ourselves to nearest-neighbor tunneling, we find $H=H_{\rm tun}+H_{\rm loc}$, with tunneling and local terms,
\bea
H_{\rm tun} &=&- \frac{t}{\sqrt{2}}  \sum_{\tbf{r}} 
      \left\{ \left[ b_x ^\dag (\tbf{r}) + b_y ^\dag (\tbf{r}) \right]
	      \left[ b_s (\tbf{r}_1)  - b_s (\tbf{r}_2)  \right] \right.
      \nonumber \\
  &&	 \left. + \left[ b_y^\dag (\tbf{r}) -b_x ^\dag (\tbf{r}) \right] 
	    \left[ b_s (\tbf{r}_3 ) -b_s (\tbf{r}_4 )  \right] 
	 +h.c.   \right\} \\
H_{\rm loc}  &=& - \sum_{\tbf{r}} \left[\mu_p n_p (\tbf{r}) + \mu_s n_s (\tbf{r}_1) \right] \nonumber \\
 &+&  \sum_\tbf{r} \frac{U_p}{2} 
    \left \{ n_p(\tbf{r}) \left[ n_p(\tbf{r}) -\frac{2}{3} \right] -\frac{1}{3} {\cal L}_z ^2 (\tbf{r}) \right\} \nonumber \\
 &+& \sum_\tbf{r} \frac{U_s}{2} n_s (\tbf{r}_1 ) 
		      \left[n_s (\tbf{r}_1 ) -1 \right].
\label{eq:HamHemmerich}
\eea
Here, $b_x(\tbf{x}) $,  $b_y (\tbf{x}) $ and $b_s (\tbf{x})$ are bosonic annihilation operators of $p_x$, $p_y$ and $s$ orbitals at 
site $\tbf{x}$. 
The position vectors $\tbf{r} = r_x \hat{a}_x + r_y \hat{a}_y$, 
with integers $r_x$ and $r_y$. 
The vector $\hat{a}_x$ ($\hat{a}_y$) is the  primitive vector of the square lattice 
in the $x$ ($y$) direction (Fig.~\ref{fig:latticephasediag}). 
The positions of $s$ orbitals are 
$\tbf{r}_1 = \tbf{r} + \frac{\hat{a}_x + \hat{a}_y}{2}$, 
$\tbf{r}_2 = \tbf{r} -\frac{\hat{a}_x + \hat{a}_y}{2} $, 
$ \tbf{r}_3 = \tbf{r}  -\frac{\hat{a}_x - \hat{a}_y}{2}$, 
and 
$\tbf{r}_4 = \tbf{r} + \frac{\hat{a}_x - \hat{a}_y}{2}$. 
 The density operators are defined as $n_p = b_x^\dag b_x + b_y^\dag b_y$ and 
$n_s = b_s ^\dag b_s$. The angular momentum operator is 
\be
{\cal L}_z = i(b_x ^\dag b_y -b_y ^\dag b_x).
\ee 
Here, we have assumed square lattice  $C_4$ rotational symmetry. 

With an analysis of the time-of-flight momentum distribution, the researchers
in~\cite{2011_Wirth_pband, 2013_Oelschlaeger_pband} found evidence suggesting a staggered $p_x\pm i p_y$ SF. 
{However, a direct measurement of its key property} --- the angular momentum order --- remains a
challenge.  This is especially crucial in the absence of superfluid coherence,
since quantum or thermal fluctuations may kill superfluidity while preserving
angular momentum order.  Such fluids with spontaneously broken time-reversal
symmetry but no superfluidity, are also thought to be relevant to
Sr$_2$RuO$_4$~\cite{2003_Maeno_SRO_RMP,2012_Kallin_SRO_RPP,2012_Nandkishore_SRO_PRB},
and to the pseudogap state of the high temperature superconductors~\cite{2000_Varma_HighTc_TRS_PRB,
  2001_Chakravarty_HighTc_TRS_PRB,2006_Fauque_HighTc_TRS_PRL}.

\ncsection{Results}

This brings us to two central results. (i) Using classical Monte Carlo
simulations of an effective model of interacting $p_x$ and $p_y$ bosons, we show
that thermal fluctuations lead to a two-step melting of the staggered $p_x \pm i
p_y$ superfluid ground state. Sandwiched between a lower temperature
Berezinskii-Kosterlitz-Thouless (BKT) transition at which superfluidity is lost,
and a higher temperature Ising transition at which time-reversal symmetry is
restored, lies a ``chiral Bose liquid'' with spontaneously broken time-reversal
symmetry. In other words, {\it it is a remarkable state of matter that is chiral but
  not superfluid.}
For large Hubbard repulsion at integer fillings, $n \geq 2$,
a strong coupling expansion yields Mott insulating states with staggered $p_x
\pm i p_y$-order. As shown schematically in Fig.~\ref{fig:latticephasediag}, this
opens up a wide window in the phase diagram where staggered angular momentum
order persists robustly even in the absence of superfluidity.

(ii) Mapping the $p_x,p_y$ orbitals onto an effective pseudospin-$1/2$
degree of freedom, we show that one can simulate the spin dynamics in
magnetic solids by orbital dynamics of $p$-band bosons.  Specifically, we
numerically study a particular lattice quench, using time-dependent matrix
product and Gutzwiller states, which is shown to convert the angular momentum
order of such chiral fluids into time-dependent oscillations of the orbital
population imbalance, analogous to Larmor spin precession. These
oscillations directly reveal the experimentally hard-to-detect ``hidden order" associated with spontaneous
time-reversal symmetry breaking. This quench is analogous to nuclear magnetic
resonance schemes in liquids or solids, which tip the nuclear moment vector and
study its subsequent precession using radio-frequency probes.
This non-interferometric route to measuring the angular momentum order works in superfluid 
as well as non-superfluid regimes, and it could be
implemented using recent experimental innovations
~\cite{2007_Bloch_superlattice_Nature, 2011_Esslinger_OrbFermionDyn_Nature, 2012_Hemmerich_TopAvoidCross_PRL}.

\paragraph{\bf Strong Coupling: $p_x \pm i p_y$ Mott Insulator and Chiral Bose Liquid.}
We begin with the strong coupling regime, where atoms can localize to form a
Mott insulator (MI) ground state. When the $s$-orbitals in one of the
  sublattices (Fig.~\ref{fig:latticephasediag}) are largely mismatched 
  in energy with 
the $p$-orbitals,
with a gap $\Delta_\mathrm{sp}$, the nearest-neighbor
tunneling Hamiltonian for bosons dominantly residing in the $p$-orbitals is
given by
\bea 
\textstyle
H_{\rm tun}^ \mathrm{eff}  &=& \sum_{\tbf{r}}\left\{  t_\parallel 
\left [ b_x ^\dag (\tbf{r})  b_x (\tbf{r} + \hat{a}_x) + 
	x \leftrightarrow y
\right] \right.\nn \\
&-&\textstyle \left. t_\perp 
\left [ b_x ^\dag (\tbf{r}) b_x (\tbf{r} + \hat{a}_y) + 
	x \leftrightarrow y
\right]
+h.c.
\right \} , 
\eea 
with hopping amplitudes $t_\parallel \approx t_\perp \approx \frac{t^2}{\Delta_\mathrm{sp}}$ 
being mediated by the $s$-orbitals. 
At integer filling, with $n\geq 2$, a strong $p$-orbital Hubbard repulsion in Eq.~\eqref{eq:HamHemmerich} 
favors a local state with a fixed particle number with nonzero angular momentum in order to minimize the interaction 
energy~\cite{2009_Wu_pband,2011_Li_EFA,2012_Li_1Dpboson_PRL,2013_Hebert_QMCpboson_PRB}, 
leading to a Mott insulator
with a two-fold degeneracy of orbital states $p_x \pm i p_y$ at each site. 
This extensive degeneracy is lifted by virtual
boson fluctuations within
second order perturbation theory in the boson hopping amplitudes. This effect is captured,
by setting ${\cal L}_z (\tbf{r}) = \sigma_z(\tbf{r}) |{\cal L}_z (\tbf{r})|$, and deriving
an effective exchange Hamiltonian
between the Ising degrees of freedom $\sigma_z(\tbf{r})$,
\bea 
H^{\rm eff}_\mathrm{Ising}  = \sum_{\langle \tbf{r}, \tbf{r}' \rangle} {\mathcal J}
      \sigma_z (\tbf{r}) \sigma_z (\tbf{r}'), 
\eea 
where
$
{\mathcal J} \! =\! \frac{3n^2 (n+2) }{2 (n+1)} \frac{t_\parallel t_\perp }{U} \!\!>\!\! 0
$. 
{The chiral  MI ground}  state thus supports a staggered (antiferromagnetic) angular
momentum pattern, with a nonzero order parameter ${\cal L}^{\rm stag}_z
(\tbf{r})= (-1)^{r_x + r_y} {\cal L}_z (\tbf{r})$ out to arbitrarily strong coupling.
Such staggered
time-reversal symmetry broken Mott insulators, albeit for far more delicate {\it plaquette}
currents, are known to 
emerge in frustrated Bose Hubbard models without orbital degrees of freedom, but
only in an extremely small parameter window of 
interactions~\cite{2012_Dhar_ChiralMott_PRA,2013_Dhar_ChiralMott_PRB,2013_Altman_ChiralMott}.
As schematically shown
in Fig.~\ref{fig:latticephasediag}, heating this MI leads to a ``chiral Bose
liquid'' with spontaneously broken time-reversal symmetry.  It only reverts to
a conventional normal fluid above a symmetry restoring thermal phase transition
of $H^{\rm eff}_\mathrm{Ising}$ which occurs at $k_B T_I \approx 2.27 {\mathcal
  J}$~\cite{1944_Onsager_IsingModel}.

\paragraph{\bf Weak Coupling: Monte Carlo Simulations.}
At weak coupling, we begin with the Hamiltonian $H^{\rm eff}_{\rm tun}$, supplemented with
local $p$-orbital interactions
\bea
H^{\rm eff}_{\rm loc} &=&   
\sum_\tbf{r} \frac{U_p}{2}  \left \{ n_p(\tbf{r}) \left[ n_p(\tbf{r}) -\frac{2}{3} \right] -\frac{1}{3} {\cal L}_z ^2 (\tbf{r}) \right\} \nonumber \\
&-& \sum_{\tbf{r}} \mu_p n_p (\tbf{r}).
\eea
For small $U_p \ll t_\parallel,t_\perp$, the band structure of $p$-band bosons has minima at $(\pi,0)$ and
$(0,\pi)$. Interactions scatter boson pairs from one minimum into the other, leading the bosons to
condense into a superposition state of the two modes, phase-locked with a relative
phase $\pm \pi/2$.
This gives rise to a $p_x \pm ip_y$ superfluid ground state with a spontaneously broken time-reversal symmetry and 
nonzero staggered angular momentum order.

\begin{figure}[htp]
\includegraphics[angle=0,width=\linewidth]{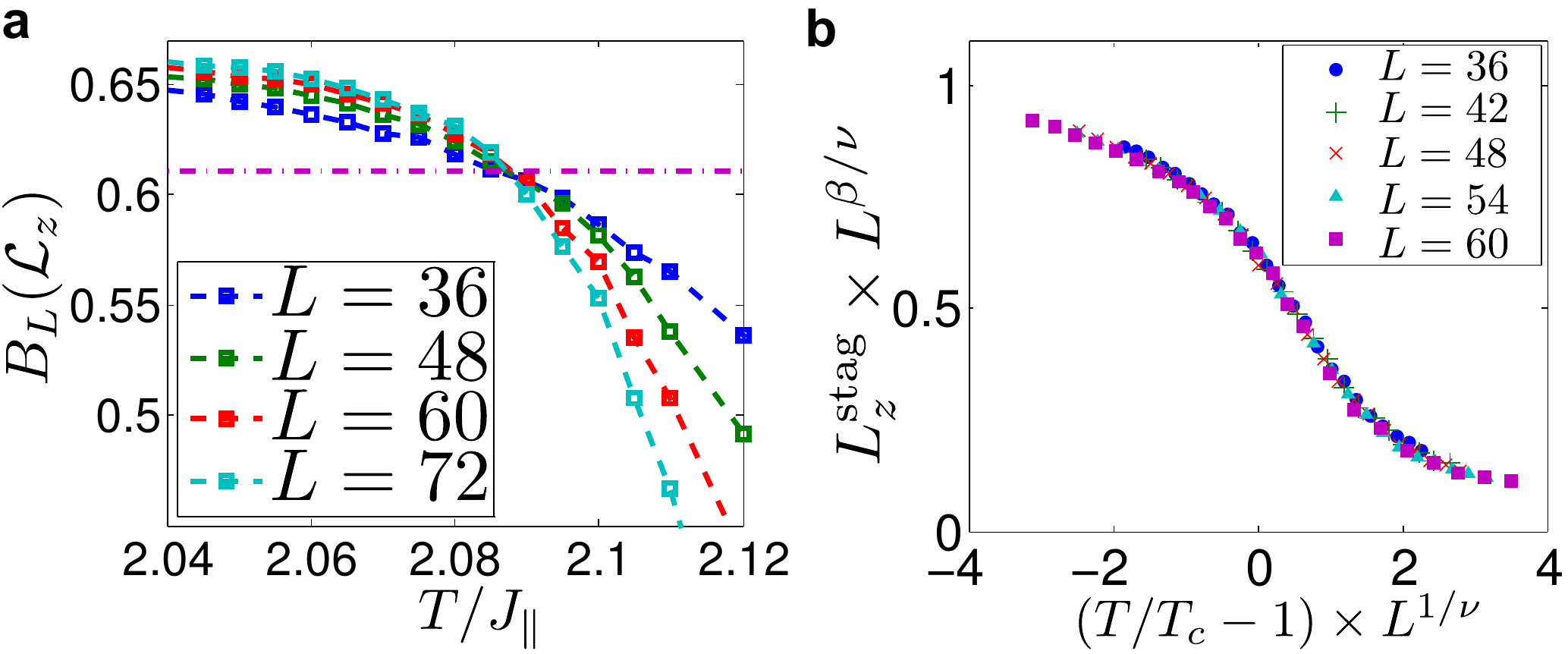}
\caption{Monte Carlo simulation results for the angular momentum ordering of the $p_x\pm i p_y$ superfluid.
(\textbf{a}), Binder cumulant $B_L({\cal L}_z)$ of the staggered angular momentum order parameter for different
system sizes $L$ showing a crossing point at the Ising transition at $T/J_\parallel = 2.088(3)$. 
The dashed line
indicates the critical Binder cumulant $0.61069\ldots$ for a 2D Ising transition.
(\textbf{b}), scaling
collapse of the angular momentum order parameter curves for Ising exponents $\nu=1$ and $\beta=1/8$.
}
\label{fig:mcBC}
\end{figure}

\begin{figure}[htp]
\includegraphics[angle=0,width=\linewidth]{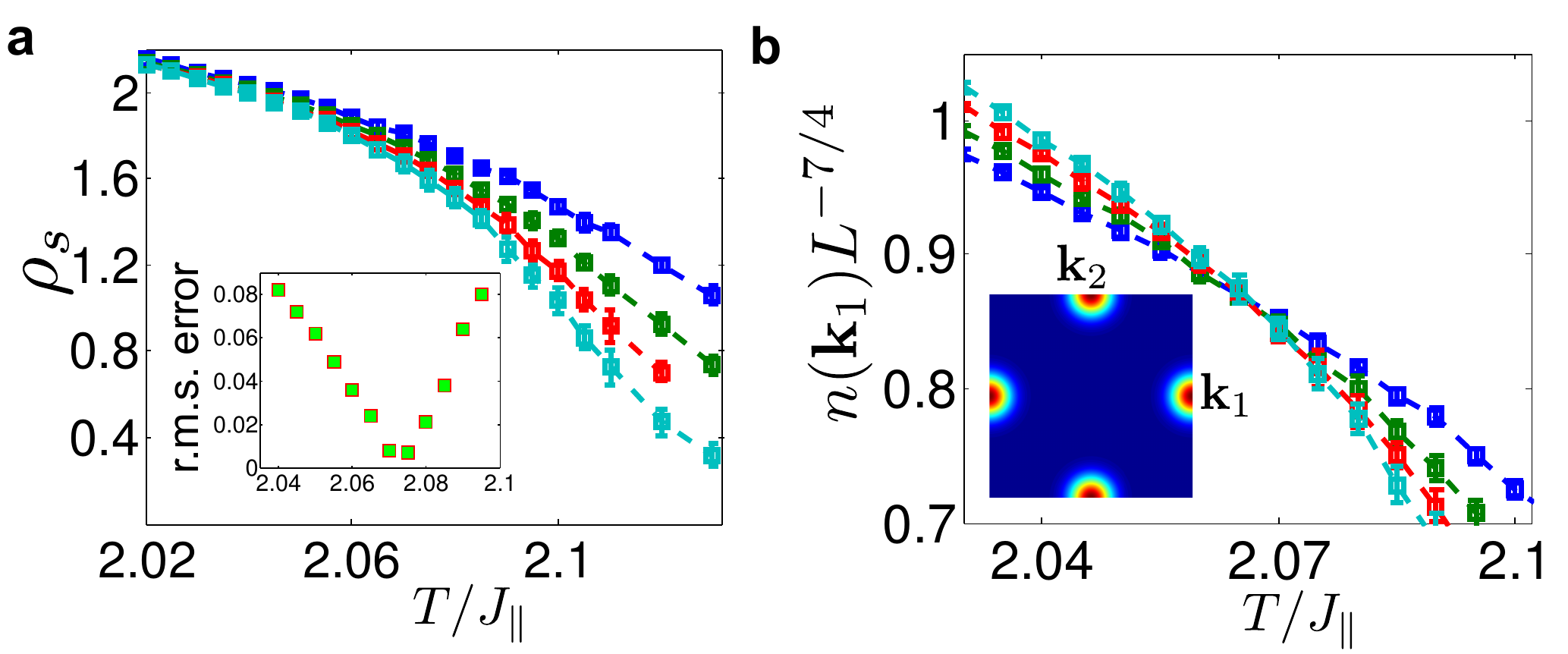}
\caption{Monte Carlo simulation results for the superfluidity of $p$-orbital bosons.
(\textbf{a}), temperature dependence of the superfluid stiffness $\rho_s$ for different system sizes $L$, showing a 
rapid drop consistent with finite size effects at a BKT transition. Inset shows the r.m.s. error from fitting $\rho_s(L)$
to the Weber-Minnhagen log-scaling form at different temperatures, with a steep minimum
at the BKT transition point $T/J_\parallel=2.072(3)$. (\textbf{b}), scaled momentum distribution $n(\tbf{k}_1) L^{-7/4}$
for different system sizes $L$, showing a crossing at the BKT transition point. Inset shows the
schematic momentum distribution over the Brillouin zone, with equal height peaks at $\tbf{k}_1$ and $\tbf{k}_2$.}
\label{fig:mcsf}
\end{figure}

To study the impact of thermal fluctuations on this weakly correlated superfluid, we make the reasonable assumption 
that classical phase fluctuations dominate the universal physics 
in the vicinity of the thermal phase transitions of this superfluid. This allows us to ignore the
subdominant density fluctuations, and to replace $b^\dagger_{x,y} \sim \sqrt{\rho/2} {\rm e}^{i \theta_{x,y}}$, with
$\rho$ being the boson density, arriving at an effective classical phase-only Hamiltonian
\bea
H^{\rm eff}_{\rm phase} \!\!\!\!&=&\!\!\!\! 
 \sum_{\tbf{r}} \left [  \left\{ 2 J_\parallel 
\cos (\Delta_x \theta_x(\tbf{r})) - 2 J_\perp 
\cos(\Delta_y \theta_x (\tbf{r}))
\right\} \right.  \\
&+& \left. \left\{ x \leftrightarrow y \right\} \right]
 - U \sum_{\tbf{r}} \sin^2(\theta_x(\tbf{r}) - \theta_y(\tbf{r}))
\label{Hphase}
\eea
where $\Delta_j \theta_\alpha(\tbf{r}) = \theta_\alpha(\tbf{r}+\hat{a}_j)-\theta_\alpha(\tbf{r})$ with $j=x,y$,
$J_{\parallel,\perp} \!  \approx \! \rho t_{\parallel,\perp}/2$ and $U \! \approx  \!  \rho^2 U_p/6$.

Using Monte Carlo simulations (see Methods), we have studied the thermal phase diagram of this model for fixed values 
of $U/J_\parallel$, with $J_\perp/J_\parallel=1$. As shown in Fig.~\ref{fig:mcBC}, the Binder cumulant \cite{1981_Binder}
$B_L({\cal L}_z)$ for the
staggered angular momentum order, computed for $U/J_\parallel=1$ on $L\times L$ systems with various $L$, exhibits a
unique crossing point at $T/J_\parallel = 2.088(3)$, signaling a critical point with a diverging correlation length. The critical value of this 
Binder cumulant is $B^* \approx 0.61$, very close to the universal 2D Ising value $\approx 0.61069$ for the aspect ratio of unity
and periodic boundary 
conditions used in our simulations. This suggests that the staggered angular momentum order disappears at $T_I/J_\parallel= 2.088(3)$ 
via a thermal transition in the 2D Ising universality class. In order to track
the destruction of superfluidity,  we
have computed the superfluid stiffness $\rho_s$ (see Methods), finding evidence of a Berezinskii-Kosterlitz-Thouless type behavior rounded by finite size effects.
A finite size scaling analysis shows that a fit to the Weber-Minnhagen 
log-scaling form \cite{1988_PRB_WeberMinnhagen}, obtained from the Kosterlitz-Thouless renormalization group equations,  
yields $T_{\rm BKT}/J_\parallel = 2.072(3)$. An unbiased fit to this log-scaling form also yields 
$\rho_s(T_{\rm BKT})/T_{\rm BKT} \approx 0.64$, very close to the universal value $2/\pi$. As further confirmation of the BKT character of the
superfluid transition, we have computed the boson
momentum distribution 
$n(\tbf{k}) = \frac{1}{L^2} \sum_{\tbf{r r'} \alpha}
	{\rm e}^{i \tbf{k} \cdot (\tbf{r} - \tbf{r}')}
	\langle {\rm e}^{i \theta_{\alpha}(\tbf{r})} {\rm e}^{- i \theta_{\alpha}(\tbf{r'})} \rangle$,
finding equal
height peaks at $\tbf{k}_1=(\pi,0)$ and $\tbf{k}_2=(0,\pi)$. This is 
consistent with the weak coupling analysis which shows $p$-band
dispersion minima at these momenta. At a BKT transition, the momentum distribution is expected to scale as $\sim L^{7/4}$ (in contrast to
scaling as $\rho_c L^{2}$ for a Bose condensate with a condensate density $\rho_c$). This implies that the
scaled momentum distributions $n(\tbf{k}_{1}) L^{-7/4}$
cross at $T_{\rm BKT}$ for various system sizes $L$; we find this occurs at $T/J_\parallel \approx 2.07$, close to the
result found from the superfluid stiffness analysis (Fig.~\ref{fig:mcsf}). Our numerical study thus shows that the $p_x \pm i p_y$ superfluid
undergoes a two-step destruction: a lower temperature BKT transition at which superfluidity is lost followed by a higher
temperature Ising transition at which time reversal symmetry is restored, leading to an unconventional ``chiral Bose liquid'' 
at intermediate temperatures $2.072(3) \lesssim T \lesssim 2.088(3)$. With increasing correlations, the BKT transition temperature is
expected to get suppressed, eventually vanishing at the Mott transition (for integer fillings $n \geq 2$), while the Ising transition
remains nonzero for arbitrarily large repulsion as seen from the earlier strong coupling limit. Correlation effects thus enhance
the window where one realizes a ``chiral Bose liquid'' as shown in the schematic temperature-interaction phase diagram in Fig.~\ref{fig:latticephasediag}.

\paragraph{\bf Quantum Quench and Single-Site Orbital Dynamics.}
One can draw a fruitful analogy between the two orbital states at each site $p_{x},p_{y}$ and a pseudospin-1/2 degree of freedom $\uparrow,\downarrow$. This suggests that one can simulate spin dynamics in solid state materials by studying orbital dynamics of $p$-band bosons. As we will see, this also suggests a route to directly detecting the angular momentum order in the $p_x \pm i p_y$ superfluid and ``chiral Bose liquid'' of the type we have obtained. In our analogy, the $p_x \pm i p_y$ state corresponds to a pseudospin pointing along the $\pm \hat{y}$ direction in spin space. Applying a ``magnetic field'' along the $\hat{x}$ direction to this pseudospin should then induce Larmor precession, leading to periodic oscillations of the $z$-magnetization, corresponding to oscillations in the orbital population-imbalance ${\cal N}(p_x)-{\cal N}(p_y)$. Let us imagine we prepare the system in a certain initial state, and then suddenly quench to a state where we set $U_p=U_s=0$, turn off all hoppings so $t=0$, and turn on a ``magnetic field'' term
\bea
H_{\rm mag}  =  \sum_\tbf{r} (-1)^{r_x + r_y} 
  \lambda (\tbf{r}) \left [b_x ^\dag (\tbf{r}) b_y (\tbf{r}) + b_y ^\dag (\tbf{r}) b_x (\tbf{r}) \right]
\label{eq:quenchH}
\eea 
at time $\tau=0$; we later discuss how to realize such a term in optical lattice experiments.
The staggered sign $(-1)^{r_x + r_y}$ leads to a staggered coupling between the $p_x$- and $p_y$-orbitals. 
If initially a staggered superposition $p_x \pm e^{i \theta} p_y$ is prepared, this results in a rectification of all local Lamor precessions such that they add up to produce a macroscopic oscillation of the populations of the  $p_x$- and $p_y$-orbitals.
The $p$-orbital imbalance, $\Delta{\cal N} (\tbf{r}) = b_x ^\dag (\tbf{r}) b_x(\tbf{r}) - b_y ^\dag (\tbf{r}) b_y (\tbf{r}) $, evolves, within a Heisenberg picture, as
\bea
\frac{d \Delta \mathcal{N} (\tbf{r}, \tau)}{d \tau} = - i  [\Delta \mathcal{N} (\tbf{r}, \tau),  H_{\rm mag}]  
= -2 \lambda (\tbf{r}) {\cal L}^{\rm stag}_{z} (\tbf{r}, \tau),
\eea
where ${\cal L}_z^{\rm stag} = {\cal L}_z (-1)^{r_x+r_y}$ is the staggered angular momentum operator whose
evolution is in turn given by
\bea
\frac{d {\cal L}^{\rm stag}_{z} (\tbf{r},\tau) }{d\tau } = 2\lambda (\tbf{r}) \Delta {\cal N} (\tbf{r}, \tau).
\eea
This leads to periodic oscillations of
$\Delta N(\tbf{r},\tau) = \langle\Delta \mathcal{N} (\tbf{r},\tau) \rangle$ as
\bea
\Delta N(\tbf{r}, \tau) \!\!\!&=&\!\!\! \Delta N (\tbf{r}, 0) \cos(2\lambda(\tbf{r})\tau) \! - \! L^{\rm stag}_z (\tbf{r}, 0) \sin(2\lambda(\tbf{r})\tau) \nn \\
&\equiv& A(\tbf{r}) \cos(2\lambda(\tbf{r}) \tau + \phi(\tbf{r})).
\eea
where $\Delta N (\tbf{r}, 0)$ and $L^{\rm stag}_z (\tbf{r}, 0)$ denote the
initial orbital magnetization and staggered angular momentum,  respectively.
Neglecting possible spatial inhomogeneity in $\lambda(\tbf{r})$ and $\phi (\tbf{r})$, by focusing at the
trap center, we can set $\lambda(\tbf{r}) = \lambda$ and $\phi(\tbf{r}) = \phi$, 
and extract the initial angular momentum order from the amplitude $A(\tbf{r})$ and the phase shift $\phi$ in 
the dynamics of the averaged number difference 
$ \overline{\Delta N}(\tau) = \frac{1}{N_s} \sum_\tbf{r} \Delta N (\tbf{r},\tau)$ with $N_s$ being the number of 
lattice sites at the trap center, and $\overline{(\ldots)}$ denoting the spatial average of $(\ldots)$.
The coefficient $\lambda$ can be directly read-off from 
the oscillation period $\tau_Q  \equiv \pi/\lambda$.
We emphasize here that $\overline{\Delta N}$ suitably averaged over the entire
trap can be measured in time-of-flight  
experiments~\cite{2011_Wirth_pband}. 

For a state with nonzero staggered angular momentum order, but no initial orbital population imbalance, i.e., 
$\Delta N(\tbf{r},0)=0$, such as our chiral fluids, we expect $\overline{\Delta N}(\tau)$ to oscillate with a nonzero amplitude, and a 
phase $\pm \pi/2$ whose sign will fluctuate from realization to realization, reflecting the spontaneous nature of time-reversal symmetry breaking. The amplitude of the signal will then be a direct measure of the staggered angular momentum order parameter, vanishing in a singular manner at
the Ising phase transition which restores time-reversal symmetry. By contrast, a completely thermally disordered conventional normal fluid
would have $\overline{\Delta N}(\tau) =0$. A state with an initial orbital population imbalance but no angular
momentum order, obtained by explicitly breaking the square lattice $C_4$ symmetry in the initial Hamiltonian
as achieved in recent experiments, would exhibit oscillations with a nonsingular amplitude and a phase $\phi=0$. Finally, if spontaneous time-reversal symmetry breaking exists in a system without $C_4$ symmetry, the amplitude of $\overline{\Delta N}(\tau)$ will be nonsingular while its phase will change in a singular manner, going from $\phi = \pm \pi/2$ in a completely ordered state to $\phi = 0$ at the time-reversal
symmetry restoring phase transition. Since this quench induced orbital magnetization dynamics is
inherently a non-interferometric probe of the angular momentum order, it suggests a simple and powerful method for measuring 
time-reversal symmetry breaking in superfluid {\it as well as} non-superfluid chiral states.
Our proposal thus significantly extends the earlier proposed quench dynamics approach for probing generic current
orders~\cite{2012_Killi_CurrentQuench_PRA,2012_Killi_AnisotropicQuench_PRA}.
In the presence of superfluid order, our real space quench is analogous to the recent proposal of Cai {\it et al.,} \cite{2011_Cai_UBEC} 
which proposes to extract the relative  phase between the $p_x$ and $p_y$ orbitals by studying momentum spectra after 
applying a Raman pulse to the Bose condensate. However, our proposal differs in showing that the
angular momentum order can be probed irrespective 
of long range phase coherence or sharp momentum peaks.

\paragraph{\bf Numerical Simulations of Quench Dynamics.}
Our above analysis assumed that the quantum quench was complete, i.e., all tunnelings ($t$) and 
interactions ($U$) were entirely switched off when $H_{\rm mag}$ was switched on. We now show,
using numerical simulations, that the coherent orbital oscillations are robust even with 
small nonzero tunnelings and interactions present after the quench, i.e., for an {\it incomplete}
quench. 

Since the scheme we are proposing here directly measures the local angular
momentum order, it does not rely on the system dimensionality (beyond the
assumption of long-range order). We therefore numerically simulate the zero
temperature quench dynamics of a 1D model of $p$-orbital bosons, using both
time-dependent Gutzwiller mean field theory~\cite{2001_Seibold_TGutz_PRL} 
and time-dependent matrix
product states (tMPS), finding good agreement at both weak and strong
couplings and qualitatively similar conclusions at intermediate interaction strength. 
We then use the Gutzwiller mean field theory to also simulate the
dynamics for the 2D case relevant to current experiments.
 
\begin{figure}[htp]
\includegraphics[angle=0,width=.95\linewidth]{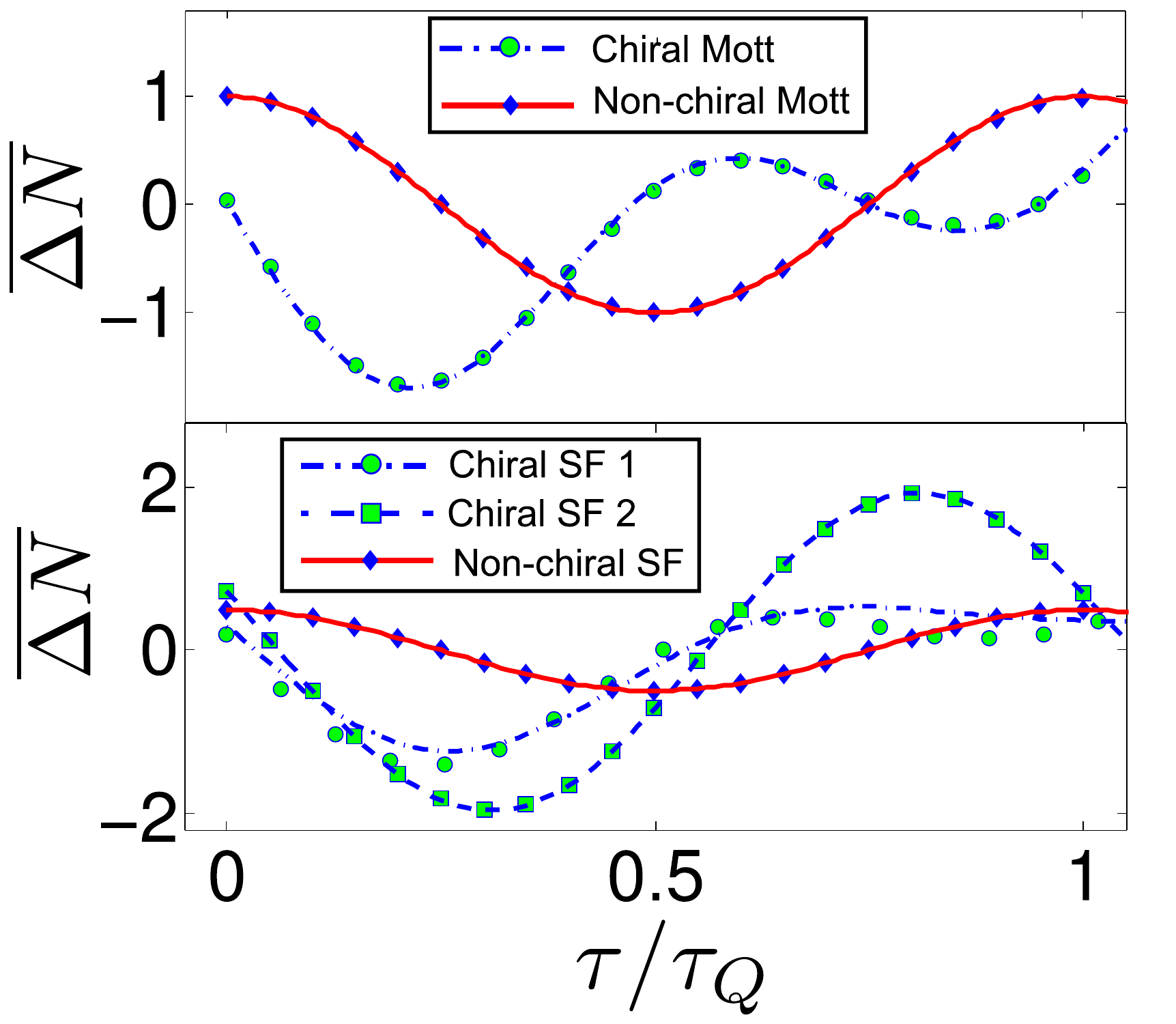}
\caption{Quench dynamics of one dimensional phases. 
Dots and lines are results of tMPS and Gutzwiller methods, respectively. 
The upper (bottom) panel shows dynamics of Mott (superfluid) states. 
The fillings for chiral and non-chiral Mott states, chiral SF~$1$, non-chiral 
SF and chiral SF~$2$ are $\langle n(\tbf{r}) \rangle =2$, $1$, $1.5$, $0.5$ and $2$, respectively.  
For the chiral SF~$2$ state, we use $t_\parallel = 2t_\perp = U/3 = \lambda/10$; while for other states 
we use $t_\parallel = 9 t_\perp = 0.045 U =0.09 \lambda$. 
The time unit $\tau_Q$ is $\pi/\lambda$.
}
\label{fig:1Dquench}
\end{figure}

The Hamiltonian of the 1D system is~\cite{2012_Li_1Dpboson_PRL}
\bea
\!\! H_{1D}\!\! &=& \!\! \sum_{j} \! \left[ t_\parallel b_x ^\dag (j) b_x(j\!+\! 1) \! - \! t_\perp b_y ^\dag (j) b_y(j\!+\! 1) \! +\! h.c.\right] \nonumber \\
\!\!&+&\!\!  \sum_j \frac{U}{2} 
    \left \{ n(j) \left[ n(j)\! -\!\frac{2}{3} \right] \! -\! \frac{1}{3} {\cal L}_z ^2 (j) \right\}, 
\eea
where $j$ index lattice sites. 
The ground state phase diagram of this 1D system includes two types of Mott 
states at strong coupling: a {chiral Mott} with staggered angular momentum order, and 
a {non-chiral Mott} insulator. For
weak correlations, it supports two types of SF ground states: 
{a chiral superfluid} with staggered angular 
momentum order and 
a {non-chiral} superfluid~\cite{2012_Li_1Dpboson_PRL}. The chiral states have an
order parameter 
${L}^{\rm stag}_z (j)  =\langle {\cal L}^{\rm stag}_z (j) \rangle$,
with
${\cal L}^{\rm stag}_z (j)   =   (-1)^j  {\cal L}_z (j) $, 
which is analogous to the 2D case.
We start with different ground states of $H_{1D}$ and study their time evolution under a quantum quench which
suddenly changes the Hamiltonian to $H_{1D} + \Delta H_{1D}$, where
\bea
\Delta H_{1D}  \!&=&\! \lambda \sum_j (-1) ^j
  \left [b_x ^\dag (j) b_y (j) + b_y ^\dag (j) b_x (j) \right].
\eea
The oscillatory dynamics of $\overline{{L}^{\rm stag}_z}$ and $\overline{\Delta N}$ is confirmed even for these 
entangled many-body states (Fig.~\ref{fig:1Dquench}). 
Since the 1D geometry does not possess $C_4$ symmetry, we expect the different
states to be distinguished by the phase shift $\phi$, not the amplitude, 
 of the oscillatory dynamics.
The chiral Mott and superfluid states develop a periodic motion 
with non-zero phase shift.
The dynamics of  non-chiral states indicate zero phase shift. 
In this way the chiral states can be distinguished from non-chiral states 
by measuring  
the phase shift, which is directly related to the angular momentum order parameter.
Deep in the chiral superfluid 
state, the phase shift is $\phi = \pm \pi/2$, and it decreases in magnitude upon 
approaching the chiral-nonchiral critical point. The phase shift 
vanishes in a singular fashion at this quantum critical point, signaling that
this phase transition associated with time-reversal symmetry can be probed by measuring the 
order parameter via the phase shift $\phi$.

\begin{figure}[htp]
\includegraphics[angle=0,width=\linewidth]{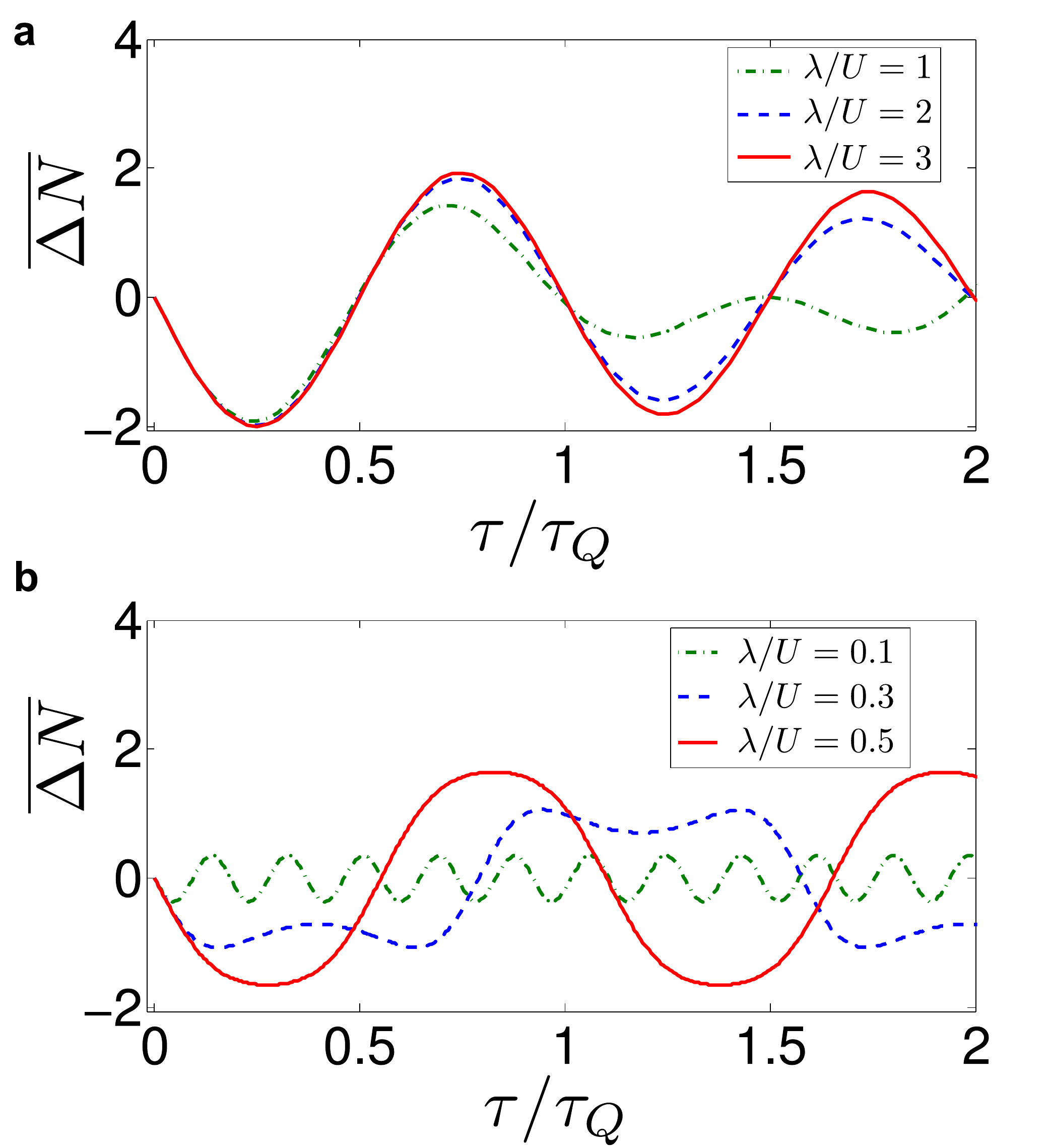}
\caption{ Quench dynamics of two dimensional phases with various quench 
strengths via Gutzwiller approach.  (a) and (b) show the evolution of $\Delta N$ 
for chiral Mott states (with $t_\parallel = t_\perp = 0.0125 U_p$) and 
superfluid states ($t_\parallel = t_\perp = U_p$), respectively.  The filling is 
$\langle n(\tbf{r}) \rangle =2$. 
The time unit $\tau_Q$ is $\pi/\lambda$. {Non-chiral
phases (not shown) have $\Delta N=0$, and do not exhibit any post-quench
oscillations.}
}
\label{fig:Gutzwillerquench}
\end{figure}

Comparing time-dependent Gutzwiller~\cite{2007_Snoek_TGutzwiller} and tMPS methods (Fig.~\ref{fig:1Dquench}), 
we find that the
Gutzwiller approach captures orbital dynamics fairly well. We thus apply this approach 
to study orbital dynamics of the two dimensional system, as in experiments~\cite{2011_Wirth_pband}. 
We have verified
that a partial quench leads to long-lived $\Delta N$ oscillations as long 
as $t$ and $U$ are weak compared with the quench strength, 
i.e., $t/\lambda \ll 1$ and $U/\lambda \ll 1$. These results are shown in Fig.~\ref{fig:Gutzwillerquench}.

\ncsection{Discussion}

In most cold atom experiments, the trap potential can induce a slowly varying inhomogeneity in the ``magnetic field'' $\lambda$ as 
$\delta \lambda = \max\{\lambda(\tbf{r})\}-\min\{\lambda(\tbf{r})\}$.
We expect the oscillations of the trap averaged number difference $\overline {\Delta N}$ to decay over a time-scale 
$\sim\frac{1}{\delta \lambda}$. 
Finally we emphasize that 
besides the angular momentum order parameter, 
the quantum quench proposal and 
a subsequent study of 
$\Delta N(\tbf{r},\tau)$ using in-situ microscopy can also yield  correlation functions of 
${\cal L}^{\rm stag}_z (\tbf{r})$, from which the diverging correlation length near the 
transition from chiral Bose liquid to normal can be extracted. 

\ncsection{Methods}

\paragraph{\bf Experimental Proposal for Quench.} 
To engineer the Hamiltonian $H_{\rm{mag}}$ of Eq.~(\ref{eq:quenchH}), we implement a quench 
potential $V_{\rm{mag}}(\tbf{x})$ {modulated in the $(1,1)$ direction }
in addition to the lattice potential giving rise to the quench Hamiltonian 
\bea 
H_{\rm{mag}} &=& 
\\ \nn
 &&\sum_\tbf{r} \left[ \epsilon(\tbf{r})  \left(b_x ^\dag (\tbf{r}) b_y (\tbf{r})  + h.c.\right )  + \mu(\tbf{r})\, n_p (\tbf{r})\right],  
\eea 
with 
\bea
\epsilon (\tbf{r}) \approx \frac{\hbar }{4  m \omega_0} 
    \frac{\partial^2 V_{\rm{mag}} \left( \tbf{r} + l \left[\frac{\hat{a}_x + \hat{a}_y }{\sqrt{2}} \right]\right) }
{a^2 \partial l^2}|_{l\to 0},
\label{eq:quenchEps}
\eea
which is valid in the tight binding regime when the quench potential is weak as 
compared to the original optical lattice (see Supplementary Information). 
Here, $\omega_0$ is the harmonic oscillator frequency of the lattice wells hosting the $p$-orbitals, 
and $a \equiv |\hat{a}_x| = |\hat{a}_y|$ is the lattice constant. Since the
local density operator of the $p$-orbitals $n_p(\tbf{r})$ commutes with $\Delta
\mathcal{N}(\tbf{r})$ and $\tilde{\cal L}_z (\tbf{r})$, it does not contribute
to the dynamics of $\Delta \mathcal{N} (\tbf{r})$, and hence may be
neglected. This is verified in our Gutzwiller simulations. {We choose}  the quench potential as
\bea 
V_{\rm{mag}}(\tbf{x} ) = - \Gamma \cos^2 \left( \frac{2 \nu+1}{4} (\hat{K}_x+\hat{K}_y) \cdot \tbf{x} \right),
\label{eq:quenchPot}
\eea
with some integer $\nu \geq 0$, a positive amplitude $\Gamma$, and $\hat{K}_x$, $\hat{K}_y$ denoting 
the primitive vectors of the reciprocal lattice ($\hat{a}_i \cdot \hat{K}_j  =  2 \pi \, \delta_{ij}$ 
with $i,j \in \{x,y\}$). The quench potential provides a lattice along the $(1,1)$ direction, and 
it  breaks both $C_4$ symmetry and mirror symmetries in the $x$ and $y$ directions. The potential 
is minimal at every second site of the Bravais lattice at positions $\tbf{r}$(=$r_x \hat{a}_x + r_y\hat{a}_y$) with even ${r_x + r_y} $ 
and it is maximal for all adjacent sites specified by  odd ${r_x + r_y}$. Hence, the second derivative 
in Eq.~(\ref{eq:quenchEps}) produces the alternating sign $(-1)^{r_x + r_y}$ required for realization 
of the quench Hamiltonian in Eq.~(\ref{eq:quenchH}). Combining Eqs.~(\ref{eq:quenchPot}) and Eq.~(\ref{eq:quenchEps}) yields 
\bea 
\epsilon (\tbf{r}) =  \frac{E_{\rm{rec}}}{\hbar \omega_0} \frac{\Gamma}{4} (2 \nu+1)^2 (-1)^{r_x + r_y},
\label{eq:quenchEps2}
\eea
with $E_{\rm{rec}} \equiv  \hbar^2 k^2 / 2 m$ denoting the single photon recoil energy for photons with wave number $k = \frac{1}{2} |\hat{K}_x+\hat{K}_y|$.

The lattice potential together with the corresponding quench potential can be realized by superlattice 
techniques demonstrated in several experiments~\cite{2007_Bloch_superlattice_Nature, 2007_NIST_superlattice_Nature, 
2011_Sengstock_honeycomb_BEC, 2011_Wirth_pband}. For example, following Ref.~\cite{2011_Wirth_pband}, 
the lattice potential arises via two optical standing waves oriented along the $(1,1)$ and $(1,-1)$ axes with a wave 
number $k = \frac{1}{2} |\hat{K}_x+\hat{K}_y| = 2 \pi / 1064$~nm. The corresponding quench lattice requires an 
additional standing wave along the $(1,1)$ axis with wave number $k' = \frac{2 \nu +1}{2} k$. Hence, 
the case $\nu=1$ requires $k' = \frac{3}{2} k \approx 2\pi / 709$~nm, which is experimentally readily provided by diode laser sources. Both standing waves along the $(1,1)$-direction may be derived by retro-reflecting two parallelly propagating laser beams with wave numbers $k$ and $k'$ by the same mirror. In order to prepare the required spatial relative phase of the two lattices, $k'$ may be slightly detuned from the precise ratio $k'/k = \frac{3}{2}$.
 
\paragraph{\bf Monte Carlo simulations.} 
We carry out the Monte Carlo study of the Hamiltonian $H^{\rm eff}_{\rm phase}$ in Eq.~(\ref{Hphase})
using a Metropolis sampling of the phase configurations $\{\theta_x(\tbf{r}),\theta_y(\tbf{r})\}$, 
with $10^{7}$ sweeps to equilibrate the system at each temperature, and averaging all observables over 
$10^{8}$ configurations. To study the phase diagram of $H^{\rm eff}_{\rm phase}$, we focus on the
angular momentum order parameter, the superfluid stiffness, and the momentum distribution, all of which
are discussed below.

The angular momentum order parameter in the phase-only effective model takes the
form of
$
{\cal M} = \sum_\tbf{r} (-1)^{r_x+r_y} \sin(\theta_x(\tbf{r}) - \theta_y(\tbf{r}))
$
and compute its Binder cumulant  \cite{1981_Binder}
$
{\cal B}_L({\cal L}_z) = 1 -  \frac{\langle {\cal M}^4 \rangle}{3 \langle {\cal M}^2 \rangle^2}.
$
The universal order parameter distribution at renormalization group fixed points leads to
universal values of ${\cal B}_{L \to \infty}$;  on finite size systems, this yields Binder cumulant
curves which cross at the critical point associated with angular momentum ordering. The critical value $B^*$ of the
Binder cumulant is well-known to be universal, independent of lattice structure and 
details of the Hamiltonian, and depending only on the aspect ratio and boundary conditions
used in the simulations. For periodic boundary conditions on $L\times L$ lattices,
$B^* \approx 0.61069$ for the 2D Ising universality class.

The superfluid stiffness $\rho_s$ is defined as the change in the free energy density in response
to a boundary condition twist; for $H^{\rm eff}_{\rm phase}$, it is explicitly given by 
\bea
\!\!\!\! \rho_s(T)  \!\!\!\!&=&\!\!\!\! \frac{1}{N} ( \langle {\cal K}_x \rangle - \frac{1}{T} \langle {\cal I}^2_x \rangle) \\
\!\!\!\!  {\cal K}_x  \!\!\!\!&=&\!\!\!\! - 2 J_\parallel \!  \sum_{\tbf{r}} \!    \cos(\Delta_x \theta_x(\tbf{r})) \!   + \!   2 J_\perp \! \!  \sum_{\tbf{r}} \!   \cos(\Delta_x \theta_y(\tbf{r})) \\
\!\!\!\!  {\cal I}_x \!\!\!\!&=&\!\!\!\!  2 J_\parallel \!   \sum_\tbf{r} \!   \sin(\Delta_x \theta_x(\tbf{r}))
\!   - \!  2 J_\perp \! \sum_\tbf{r} \sin(\Delta_x \theta_y(\tbf{r}))
\eea
where $\langle \cdots \rangle$ refers to the thermal average.
At a BKT transition, $\rho_s(T)$ jumps to zero, with $\rho_s(T_{\rm BKT})/T_{\rm BKT}=2/\pi$, a 
universal value.
On finite size systems, the
universal superfluid stiffness jump gets severely rounded, and a careful 
finite size scaling is required to extract $T_{\rm BKT}$. Based on the KT renormalization group equations, 
Weber and Minnhagen have shown \cite{1988_PRB_WeberMinnhagen} that $\rho_s(T_{\rm BKT},L)$ scales as
\bea
\rho_s(T_{\rm BKT},L) = \rho_s(T_{\rm BKT},\infty)(1+\frac{1}{2\log L + c})
\eea
where $c$ is a non-universal number. It is well-known that fitting to this log-scaling form
at different temperatures leads to an error which exhibits a steep minimum at $T_{\rm BKT}$, 
enabling us to extract $T_{\rm BKT}$ from our simulations. An unbiased fit to $\rho_s(T)/T$, using a 
two-parameter scaling form,
\bea
\frac{\rho_s(T_{\rm BKT},L)}{T_{\rm BKT}} = a(1+\frac{1}{2\log L + c})
\eea
also enables one to confirm the universal jump at the $T_{\rm BKT}$ identified by the error minimum.
Using this, we find $a \approx 0.64$ from our simulations, in very good agreement with the KT value 
$2/\pi=0.6366...$.

\ncsection{Acknowledgment} The authors acknowledge helpful discussions
with Youjin Deng, Subroto Mukerjee and Sankar Das Sarma.  This
work is supported by the NSERC of Canada (AP), NSF PHY11-25915 (X.L.), AFOSR
(FA9550-12-1-0079), ARO (W911NF-11-1-0230), DARPA OLE Program through ARO 
and the Charles E. Kaufman Foundation of The Pittsburgh Foundation 
(X.L. and W.V.L.), the National Basic Research Program of China (Grant No
2012CB922101) and Overseas Collaboration Program of NSF of China (11128407)
(W.V.L.), and the German Research Foundation DFG-SFB 925 (A.H.). 
X.L. would like to thank KITP at UCSB for hospitality.
W.V.L. and A.H. acknowledge partial support by NSF-PHYS-1066293 and the hospitality of the Aspen Center for Physics.

\section*{Author contributions}
X.L. and A.P. conceived and evolved the theoretical ideas in discussion with
W.V.L. A.H. examined and improved the experimental protocol.  X.L. and
A.P. performed numerical simulations. All authors worked on theoretical analysis
and contributed in completing the paper.

\section*{Author Information}
The authors declare no competing financial interests.
Correspondence and requests for material should be sent to
w.vincent.liu@gmail.com.
Supplementary information accompanies this paper.

\begin{widetext}
\newpage 

\renewcommand{\thesection}{S-\arabic{section}}
\renewcommand{\theequation}{S\arabic{equation}}
\setcounter{equation}{0}  
\renewcommand{\thefigure}{S\arabic{figure}}
\setcounter{figure}{0}  

\section*{\Large\bf Supplementary Information}
\begin{center}
{\large \bf \papertitle} 
\end{center}




\section{Derivation of the quench strength} 
The induced coupling between $p_x$ and $p_y$ orbitals by the quench potential $V_{\rm{mag}}(\tbf{r})$ is 
\bea 
H_{\rm{mag}} = \sum_{\alpha \beta, \tbf{r}} g_{\alpha \beta} (\tbf{r}) b_\alpha ^\dag (\tbf{r}) b_\beta (\tbf{r}), 
\eea 
with 
$$
g_{\alpha \beta} (\tbf{r}) =  \int d^2 \tbf{x} \, w_\alpha ^ * ( \tbf{x} - \tbf{r}) V_{\rm{mag}}(\tbf{x})
w_\beta (\tbf{x} - \tbf{r}), 
$$
where $w_{\alpha = x/y} (\tbf{x})$ are Wannier functions for  $p_x$ and $p_y$ bands. The Wannier functions may be approximated by localized harmonic oscillator wavefunctions, with their widths determined by the harmonic oscillator frequency $\omega_0$. This approximation is valid in the tight binding regime to estimate local quantities, as $g_{\alpha \beta}$. For simplicity, calculations are done in a transformed basis defined by 
\bea 
\left[ \begin{array}{c}
        \tilde{b}_x \\ 
	 \tilde{b}_y 
       \end{array}
\right] 
= \left[ 
   \begin{array}{c}
   {b_x + b_y} \\
   {b_x - b_y} 
  \end{array} \right] /\sqrt{2} . 
\eea 
In this basis, the induced coupling reads $\tilde{g}_{\alpha \beta} \tilde{b}_\alpha ^\dag \tilde{b}_\beta $, with 
\bea 
\tilde{g}_{\alpha \beta} (\tbf{r}) 
=  \int d\tilde{x} d\tilde{y}   
    {w}_\alpha ^ * (\tilde{x}, \tilde{y}) 
   \tilde{V}_{\rm{mag}}(\tilde{x}) 
  {w}_\beta (\tilde{x}, \tilde{y}), 
\label{eq:tildegdef}
\eea 
where 
$\tilde{x} = [(x+y) - (r_x +r_y)a]/\sqrt{2}$, $\tilde{y} = [x-y -(r_x- r_y)a]/\sqrt{2}$ ,
with $a$ the lattice constant. 
And the potential $V_{\rm{mag}}(\tbf{x})$ in the transformed coordinates reads as 
$\tilde{V}_{\rm{mag}}(\tilde{x}) = -\Gamma \cos^2 [ \frac{(2m +1)k}{2}\tilde{x} ] $, 
with $k = \frac{1}{2} |\hat{K}_x+\hat{K}_y|$ and $a = \frac{\sqrt{2} \pi}{k}$ denoting the lattice constant. 
Since Wannier functions are localized, we can approximate the quench potential by 
\bea 
\tilde{V}_{\rm{mag}}(\tilde{x}) = V_{\rm{mag}}(\tbf{r}) 
+ \frac{1}{2}  \tilde{x}^2  \frac{d^2 {\tilde{V}_{\rm{mag}}} } {d \tilde{x}^2 }|_{\tilde{x} =0}. 
\label{eq:DVapp}
\eea 
The derivative term may be rewritten as 
\bea 
\frac{d^2 {\tilde{V}_{\rm{mag}}} } {d \tilde{x}^2 }|_{\tilde{x} =0} 
 =  \frac{1}{a^2 } \frac{\partial^2 V_{\rm{mag}} \left( \tbf{r} + l \frac{\hat{a}_x + \hat{a}_y}{\sqrt{2}} \right) }{\partial l^2 } |_{l = 0}. 
\eea 

From Eq.~\eqref{eq:tildegdef} and Eq.~\eqref{eq:DVapp}, we get 
\be 
[\tilde{g}] = g^{(d)} (\tbf{r})  \mathbb{I}  + \epsilon (\tbf{r}) \sigma_z,
\label{eq:tildeg}
\ee
with 
\bea 
\epsilon (\tbf{r}) = \frac{\hbar}{4 m \omega_0  } 
\frac{\partial^2 V_{\rm{mag}}\left( \tbf{r} + l \frac{\hat{a}_x + \hat{a}_y}{\sqrt{2}} \right) }{a^2\partial l^2 } |_{l = 0}, 
\eea 
and 
$\mathbb{I} = \left[ 
\begin{array}{cc}
 1 & 0 \\
 0 & 1
\end{array} \right] 
$, 
$\sigma_z = \left[ 
\begin{array}{cc} 
 1 & 0 \\
0 & -1 
\end{array}
\right] $.

Transforming back to the original basis, we get a coupling term 
\bea 
 \sum_{\tbf{r}} \epsilon(\tbf{r})\left[b_x ^\dag (\tbf{r}) b_y (\tbf{r}) + h.c. \right]. 
\eea  
In Gutzwiller simulations, the neglected diagonal part is studied and we find that 
its modification of the orbital dynamics is minor.

\end{widetext}

\bibliography{porbital}

\begin{thebibliography}{49}
\expandafter\ifx\csname natexlab\endcsname\relax\def\natexlab#1{#1}\fi
\expandafter\ifx\csname bibnamefont\endcsname\relax
  \def\bibnamefont#1{#1}\fi
\expandafter\ifx\csname bibfnamefont\endcsname\relax
  \def\bibfnamefont#1{#1}\fi
\expandafter\ifx\csname citenamefont\endcsname\relax
  \def\citenamefont#1{#1}\fi
\expandafter\ifx\csname url\endcsname\relax
  \def\url#1{\texttt{#1}}\fi
\expandafter\ifx\csname urlprefix\endcsname\relax\def\urlprefix{URL }\fi
\providecommand{\bibinfo}[2]{#2}
\providecommand{\eprint}[2][]{\url{#2}}

\bibitem[{\citenamefont{Wirth et~al.}(2011)\citenamefont{Wirth, \"Olschl\"ager,
  and Hemmerich}}]{2011_Wirth_pband}
\bibinfo{author}{\bibfnamefont{G.}~\bibnamefont{Wirth}},
  \bibinfo{author}{\bibfnamefont{M.}~\bibnamefont{\"Olschl\"ager}},
  \bibnamefont{and}
  \bibinfo{author}{\bibfnamefont{A.}~\bibnamefont{Hemmerich}},
  \bibinfo{journal}{Nature Physics} \textbf{\bibinfo{volume}{7}},
  \bibinfo{pages}{147} (\bibinfo{year}{2011}).

\bibitem[{\citenamefont{\"Olschl\"ager
  et~al.}(2013)\citenamefont{\"Olschl\"ager, Kock, Wirth, Ewerbeck, Smith, and
  Hemmerich}}]{2013_Oelschlaeger_pband}
\bibinfo{author}{\bibfnamefont{M.}~\bibnamefont{\"Olschl\"ager}},
  \bibinfo{author}{\bibfnamefont{T.}~\bibnamefont{Kock}},
  \bibinfo{author}{\bibfnamefont{G.}~\bibnamefont{Wirth}},
  \bibinfo{author}{\bibfnamefont{A.}~\bibnamefont{Ewerbeck}},
  \bibinfo{author}{\bibfnamefont{C.~M.} \bibnamefont{Smith}}, \bibnamefont{and}
  \bibinfo{author}{\bibfnamefont{A.}~\bibnamefont{Hemmerich}},
  \bibinfo{journal}{New Journal of Physics} \textbf{\bibinfo{volume}{15}},
  \bibinfo{pages}{083041} (\bibinfo{year}{2013}).

\bibitem[{\citenamefont{Li et~al.}(2011)\citenamefont{Li, Zhao, and
  Liu}}]{2011_Li_EFA}
\bibinfo{author}{\bibfnamefont{X.}~\bibnamefont{Li}},
  \bibinfo{author}{\bibfnamefont{E.}~\bibnamefont{Zhao}}, \bibnamefont{and}
  \bibinfo{author}{\bibfnamefont{W.~V.} \bibnamefont{Liu}},
  \bibinfo{journal}{Phys. Rev. A} \textbf{\bibinfo{volume}{83}},
  \bibinfo{pages}{063626} (\bibinfo{year}{2011}).

\bibitem[{\citenamefont{Onsager}(1944)}]{1944_Onsager_IsingModel}
\bibinfo{author}{\bibfnamefont{L.}~\bibnamefont{Onsager}},
  \bibinfo{journal}{Phys. Rev.} \textbf{\bibinfo{volume}{65}},
  \bibinfo{pages}{117} (\bibinfo{year}{1944}).

\bibitem[{\citenamefont{Bednorz and M\"uller}(1986)}]{1986_HighTc}
\bibinfo{author}{\bibfnamefont{J.}~\bibnamefont{Bednorz}} \bibnamefont{and}
  \bibinfo{author}{\bibfnamefont{K.}~\bibnamefont{M\"uller}},
  \bibinfo{journal}{Zeitschrift f\"ur Physik B Condensed Matter}
  \textbf{\bibinfo{volume}{64}}, \bibinfo{pages}{189} (\bibinfo{year}{1986}).

\bibitem[{\citenamefont{Kamihara et~al.}(2006)\citenamefont{Kamihara,
  Hiramatsu, Hirano, Kawamura, Yanagi, Kamiya, and
  Hosono}}]{2006_Pnictides_Kamihara}
\bibinfo{author}{\bibfnamefont{Y.}~\bibnamefont{Kamihara}},
  \bibinfo{author}{\bibfnamefont{H.}~\bibnamefont{Hiramatsu}},
  \bibinfo{author}{\bibfnamefont{M.}~\bibnamefont{Hirano}},
  \bibinfo{author}{\bibfnamefont{R.}~\bibnamefont{Kawamura}},
  \bibinfo{author}{\bibfnamefont{H.}~\bibnamefont{Yanagi}},
  \bibinfo{author}{\bibfnamefont{T.}~\bibnamefont{Kamiya}}, \bibnamefont{and}
  \bibinfo{author}{\bibfnamefont{H.}~\bibnamefont{Hosono}},
  \bibinfo{journal}{Journal of the American Chemical Society}
  \textbf{\bibinfo{volume}{128}}, \bibinfo{pages}{10012}
  (\bibinfo{year}{2006}), \bibinfo{note}{pMID: 16881620}.

\bibitem[{\citenamefont{von Helmolt et~al.}(1993)\citenamefont{von Helmolt,
  Wecker, Holzapfel, Schultz, and Samwer}}]{1993_CMR_vonHelmolt}
\bibinfo{author}{\bibfnamefont{R.}~\bibnamefont{von Helmolt}},
  \bibinfo{author}{\bibfnamefont{J.}~\bibnamefont{Wecker}},
  \bibinfo{author}{\bibfnamefont{B.}~\bibnamefont{Holzapfel}},
  \bibinfo{author}{\bibfnamefont{L.}~\bibnamefont{Schultz}}, \bibnamefont{and}
  \bibinfo{author}{\bibfnamefont{K.}~\bibnamefont{Samwer}},
  \bibinfo{journal}{Phys. Rev. Lett.} \textbf{\bibinfo{volume}{71}},
  \bibinfo{pages}{2331} (\bibinfo{year}{1993}).

\bibitem[{\citenamefont{{Luke} et~al.}(1998)\citenamefont{{Luke}, {Fudamoto},
  {Kojima}, {Larkin}, {Merrin}, {Nachumi}, {Uemura}, {Maeno}, {Mao}, {Mori}
  et~al.}}]{1998_Sr214_Maeno}
\bibinfo{author}{\bibfnamefont{G.~M.} \bibnamefont{{Luke}}},
  \bibinfo{author}{\bibfnamefont{Y.}~\bibnamefont{{Fudamoto}}},
  \bibinfo{author}{\bibfnamefont{K.~M.} \bibnamefont{{Kojima}}},
  \bibinfo{author}{\bibfnamefont{M.~I.} \bibnamefont{{Larkin}}},
  \bibinfo{author}{\bibfnamefont{J.}~\bibnamefont{{Merrin}}},
  \bibinfo{author}{\bibfnamefont{B.}~\bibnamefont{{Nachumi}}},
  \bibinfo{author}{\bibfnamefont{Y.~J.} \bibnamefont{{Uemura}}},
  \bibinfo{author}{\bibfnamefont{Y.}~\bibnamefont{{Maeno}}},
  \bibinfo{author}{\bibfnamefont{Z.~Q.} \bibnamefont{{Mao}}},
  \bibinfo{author}{\bibfnamefont{Y.}~\bibnamefont{{Mori}}},
  \bibnamefont{et~al.}, \bibinfo{journal}{\nat} \textbf{\bibinfo{volume}{394}},
  \bibinfo{pages}{558} (\bibinfo{year}{1998}).

\bibitem[{\citenamefont{Tokura and
  Nagaosa}(2000)}]{2000_Tokura_Nagaosa_orbital_Science}
\bibinfo{author}{\bibfnamefont{Y.}~\bibnamefont{Tokura}} \bibnamefont{and}
  \bibinfo{author}{\bibfnamefont{N.}~\bibnamefont{Nagaosa}},
  \bibinfo{journal}{Science} \textbf{\bibinfo{volume}{288}},
  \bibinfo{pages}{462} (\bibinfo{year}{2000}).

\bibitem[{\citenamefont{Lewenstein and
  Liu}(2011)}]{2011_Lewenstein_Liu_orbital_dance}
\bibinfo{author}{\bibfnamefont{M.~A.} \bibnamefont{Lewenstein}}
  \bibnamefont{and} \bibinfo{author}{\bibfnamefont{W.~V.} \bibnamefont{Liu}},
  \bibinfo{journal}{Nature Physics} \textbf{\bibinfo{volume}{7}},
  \bibinfo{pages}{101} (\bibinfo{year}{2011}).

\bibitem[{\citenamefont{Zhao and Liu}(2008)}]{2008_Zhao_pmott}
\bibinfo{author}{\bibfnamefont{E.}~\bibnamefont{Zhao}} \bibnamefont{and}
  \bibinfo{author}{\bibfnamefont{W.~V.} \bibnamefont{Liu}},
  \bibinfo{journal}{Phys. Rev. Lett.} \textbf{\bibinfo{volume}{100}},
  \bibinfo{pages}{160403} (\bibinfo{year}{2008}).

\bibitem[{\citenamefont{Zhang et~al.}(2010)\citenamefont{Zhang, Hung, Ho, Zhao,
  and Liu}}]{2010_Zhang_sppair}
\bibinfo{author}{\bibfnamefont{Z.}~\bibnamefont{Zhang}},
  \bibinfo{author}{\bibfnamefont{H.-H.} \bibnamefont{Hung}},
  \bibinfo{author}{\bibfnamefont{C.~M.} \bibnamefont{Ho}},
  \bibinfo{author}{\bibfnamefont{E.}~\bibnamefont{Zhao}}, \bibnamefont{and}
  \bibinfo{author}{\bibfnamefont{W.~V.} \bibnamefont{Liu}},
  \bibinfo{journal}{Phys. Rev. A} \textbf{\bibinfo{volume}{82}},
  \bibinfo{pages}{033610} (\bibinfo{year}{2010}).

\bibitem[{\citenamefont{Cai et~al.}(2011)\citenamefont{Cai, Wang, and
  Wu}}]{2010_Cai_FFLO}
\bibinfo{author}{\bibfnamefont{Z.}~\bibnamefont{Cai}},
  \bibinfo{author}{\bibfnamefont{Y.}~\bibnamefont{Wang}}, \bibnamefont{and}
  \bibinfo{author}{\bibfnamefont{C.}~\bibnamefont{Wu}}, \bibinfo{journal}{Phys.
  Rev. A} \textbf{\bibinfo{volume}{83}}, \bibinfo{pages}{063621}
  (\bibinfo{year}{2011}).

\bibitem[{\citenamefont{Hung et~al.}(2011)\citenamefont{Hung, Lee, and
  Wu}}]{2009_Hung_fpair}
\bibinfo{author}{\bibfnamefont{H.-H.} \bibnamefont{Hung}},
  \bibinfo{author}{\bibfnamefont{W.-C.} \bibnamefont{Lee}}, \bibnamefont{and}
  \bibinfo{author}{\bibfnamefont{C.}~\bibnamefont{Wu}}, \bibinfo{journal}{Phys.
  Rev. B} \textbf{\bibinfo{volume}{83}}, \bibinfo{pages}{144506}
  (\bibinfo{year}{2011}).

\bibitem[{\citenamefont{Zhang et~al.}(2012)\citenamefont{Zhang, Li, and
  Liu}}]{2011_Zhang_pdw}
\bibinfo{author}{\bibfnamefont{Z.}~\bibnamefont{Zhang}},
  \bibinfo{author}{\bibfnamefont{X.}~\bibnamefont{Li}}, \bibnamefont{and}
  \bibinfo{author}{\bibfnamefont{W.~V.} \bibnamefont{Liu}},
  \bibinfo{journal}{Phys. Rev. A} \textbf{\bibinfo{volume}{85}},
  \bibinfo{pages}{053606} (\bibinfo{year}{2012}).

\bibitem[{\citenamefont{Sun et~al.}(2012)\citenamefont{Sun, Liu, Hemmerich, and
  Das~Sarma}}]{2011_Sun_TSM}
\bibinfo{author}{\bibfnamefont{K.}~\bibnamefont{Sun}},
  \bibinfo{author}{\bibfnamefont{W.~V.} \bibnamefont{Liu}},
  \bibinfo{author}{\bibfnamefont{A.}~\bibnamefont{Hemmerich}},
  \bibnamefont{and}
  \bibinfo{author}{\bibfnamefont{S.}~\bibnamefont{Das~Sarma}},
  \bibinfo{journal}{Nature Physics} \textbf{\bibinfo{volume}{8}},
  \bibinfo{pages}{67} (\bibinfo{year}{2012}).

\bibitem[{\citenamefont{{Li} et~al.}(2013)\citenamefont{{Li}, {Zhao}, and
  {Liu}}}]{2012_Li_orbitalladder_NatComm}
\bibinfo{author}{\bibfnamefont{X.}~\bibnamefont{{Li}}},
  \bibinfo{author}{\bibfnamefont{E.}~\bibnamefont{{Zhao}}}, \bibnamefont{and}
  \bibinfo{author}{\bibfnamefont{W.~V.} \bibnamefont{{Liu}}},
  \bibinfo{journal}{Nature Commun} \textbf{\bibinfo{volume}{4}},
  \bibinfo{pages}{1523} (\bibinfo{year}{2013}).

\bibitem[{\citenamefont{Isacsson and Girvin}(2005)}]{2005_Isacsson_pband}
\bibinfo{author}{\bibfnamefont{A.}~\bibnamefont{Isacsson}} \bibnamefont{and}
  \bibinfo{author}{\bibfnamefont{S.~M.} \bibnamefont{Girvin}},
  \bibinfo{journal}{Phys. Rev. A} \textbf{\bibinfo{volume}{72}},
  \bibinfo{pages}{053604} (\bibinfo{year}{2005}).

\bibitem[{\citenamefont{Liu and Wu}(2006)}]{2006_Liu_TSOC}
\bibinfo{author}{\bibfnamefont{W.~V.} \bibnamefont{Liu}} \bibnamefont{and}
  \bibinfo{author}{\bibfnamefont{C.}~\bibnamefont{Wu}}, \bibinfo{journal}{Phys.
  Rev. A} \textbf{\bibinfo{volume}{74}}, \bibinfo{pages}{013607}
  (\bibinfo{year}{2006}).

\bibitem[{\citenamefont{Kuklov}(2006)}]{2006_Kuklov_sf}
\bibinfo{author}{\bibfnamefont{A.~B.} \bibnamefont{Kuklov}},
  \bibinfo{journal}{Phys. Rev. Lett.} \textbf{\bibinfo{volume}{97}},
  \bibinfo{pages}{110405} (\bibinfo{year}{2006}).

\bibitem[{\citenamefont{Lim et~al.}(2008)\citenamefont{Lim, Smith, and
  Hemmerich}}]{2008_Lim_TSOC}
\bibinfo{author}{\bibfnamefont{L.-K.} \bibnamefont{Lim}},
  \bibinfo{author}{\bibfnamefont{C.~M.} \bibnamefont{Smith}}, \bibnamefont{and}
  \bibinfo{author}{\bibfnamefont{A.}~\bibnamefont{Hemmerich}},
  \bibinfo{journal}{Phys. Rev. Lett.} \textbf{\bibinfo{volume}{100}},
  \bibinfo{pages}{130402} (\bibinfo{year}{2008}).

\bibitem[{\citenamefont{Stojanovi\ifmmode~\acute{c}\else \'{c}\fi{}
  et~al.}(2008)\citenamefont{Stojanovi\ifmmode~\acute{c}\else \'{c}\fi{}, Wu,
  Liu, and Das~Sarma}}]{2008_Vladimir_icsf}
\bibinfo{author}{\bibfnamefont{V.~M.}
  \bibnamefont{Stojanovi\ifmmode~\acute{c}\else \'{c}\fi{}}},
  \bibinfo{author}{\bibfnamefont{C.}~\bibnamefont{Wu}},
  \bibinfo{author}{\bibfnamefont{W.~V.} \bibnamefont{Liu}}, \bibnamefont{and}
  \bibinfo{author}{\bibfnamefont{S.}~\bibnamefont{Das~Sarma}},
  \bibinfo{journal}{Phys. Rev. Lett.} \textbf{\bibinfo{volume}{101}},
  \bibinfo{pages}{125301} (\bibinfo{year}{2008}).

\bibitem[{\citenamefont{Zhou et~al.}(2011)\citenamefont{Zhou, Porto, and
  Das~Sarma}}]{2010_Zhou_interband}
\bibinfo{author}{\bibfnamefont{Q.}~\bibnamefont{Zhou}},
  \bibinfo{author}{\bibfnamefont{J.~V.} \bibnamefont{Porto}}, \bibnamefont{and}
  \bibinfo{author}{\bibfnamefont{S.}~\bibnamefont{Das~Sarma}},
  \bibinfo{journal}{Phys. Rev. B} \textbf{\bibinfo{volume}{83}},
  \bibinfo{pages}{195106} (\bibinfo{year}{2011}).

\bibitem[{\citenamefont{Soltan-Panahi et~al.}(2012)\citenamefont{Soltan-Panahi,
  L\"uhmann, Struck, Windpassinger, and
  Sengstock}}]{2011_Sengstock_honeycomb_BEC}
\bibinfo{author}{\bibfnamefont{P.}~\bibnamefont{Soltan-Panahi}},
  \bibinfo{author}{\bibfnamefont{D.-S.} \bibnamefont{L\"uhmann}},
  \bibinfo{author}{\bibfnamefont{J.}~\bibnamefont{Struck}},
  \bibinfo{author}{\bibfnamefont{P.}~\bibnamefont{Windpassinger}},
  \bibnamefont{and}
  \bibinfo{author}{\bibfnamefont{K.}~\bibnamefont{Sengstock}},
  \bibinfo{journal}{Nature Physics} \textbf{\bibinfo{volume}{8}},
  \bibinfo{pages}{71} (\bibinfo{year}{2012}).

\bibitem[{\citenamefont{Cai and Wu}(2011)}]{2011_Cai_TSOC}
\bibinfo{author}{\bibfnamefont{Z.}~\bibnamefont{Cai}} \bibnamefont{and}
  \bibinfo{author}{\bibfnamefont{C.}~\bibnamefont{Wu}}, \bibinfo{journal}{Phys.
  Rev. A} \textbf{\bibinfo{volume}{84}}, \bibinfo{pages}{033635}
  (\bibinfo{year}{2011}).

\bibitem[{\citenamefont{Li et~al.}(2012)\citenamefont{Li, Zhang, and
  Liu}}]{2012_Li_1Dpboson_PRL}
\bibinfo{author}{\bibfnamefont{X.}~\bibnamefont{Li}},
  \bibinfo{author}{\bibfnamefont{Z.}~\bibnamefont{Zhang}}, \bibnamefont{and}
  \bibinfo{author}{\bibfnamefont{W.~V.} \bibnamefont{Liu}},
  \bibinfo{journal}{Phys. Rev. Lett.} \textbf{\bibinfo{volume}{108}},
  \bibinfo{pages}{175302} (\bibinfo{year}{2012}).

\bibitem[{\citenamefont{H\'ebert et~al.}(2013)\citenamefont{H\'ebert, Cai,
  Rousseau, Wu, Scalettar, and Batrouni}}]{2013_Hebert_QMCpboson_PRB}
\bibinfo{author}{\bibfnamefont{F.}~\bibnamefont{H\'ebert}},
  \bibinfo{author}{\bibfnamefont{Z.}~\bibnamefont{Cai}},
  \bibinfo{author}{\bibfnamefont{V.~G.} \bibnamefont{Rousseau}},
  \bibinfo{author}{\bibfnamefont{C.}~\bibnamefont{Wu}},
  \bibinfo{author}{\bibfnamefont{R.~T.} \bibnamefont{Scalettar}},
  \bibnamefont{and} \bibinfo{author}{\bibfnamefont{G.~G.}
  \bibnamefont{Batrouni}}, \bibinfo{journal}{Phys. Rev. B}
  \textbf{\bibinfo{volume}{87}}, \bibinfo{pages}{224505}
  (\bibinfo{year}{2013}).

\bibitem[{\citenamefont{Hauke et~al.}(2011)\citenamefont{Hauke, Zhao, Goyal,
  Deutsch, Liu, and Lewenstein}}]{2011_Hauke_OrbFermion_PRA}
\bibinfo{author}{\bibfnamefont{P.}~\bibnamefont{Hauke}},
  \bibinfo{author}{\bibfnamefont{E.}~\bibnamefont{Zhao}},
  \bibinfo{author}{\bibfnamefont{K.}~\bibnamefont{Goyal}},
  \bibinfo{author}{\bibfnamefont{I.~H.} \bibnamefont{Deutsch}},
  \bibinfo{author}{\bibfnamefont{W.~V.} \bibnamefont{Liu}}, \bibnamefont{and}
  \bibinfo{author}{\bibfnamefont{M.}~\bibnamefont{Lewenstein}},
  \bibinfo{journal}{Phys. Rev. A} \textbf{\bibinfo{volume}{84}},
  \bibinfo{pages}{051603} (\bibinfo{year}{2011}).

\bibitem[{\citenamefont{Wu}(2009)}]{2009_Wu_pband}
\bibinfo{author}{\bibfnamefont{C.}~\bibnamefont{Wu}}, \bibinfo{journal}{Mod.
  Phys. Lett. B} \textbf{\bibinfo{volume}{23}}, \bibinfo{pages}{1}
  (\bibinfo{year}{2009}).

\bibitem[{\citenamefont{Mackenzie and Maeno}(2003)}]{2003_Maeno_SRO_RMP}
\bibinfo{author}{\bibfnamefont{A.~P.} \bibnamefont{Mackenzie}}
  \bibnamefont{and} \bibinfo{author}{\bibfnamefont{Y.}~\bibnamefont{Maeno}},
  \bibinfo{journal}{Rev. Mod. Phys.} \textbf{\bibinfo{volume}{75}},
  \bibinfo{pages}{657} (\bibinfo{year}{2003}).

\bibitem[{\citenamefont{Kallin}(2012)}]{2012_Kallin_SRO_RPP}
\bibinfo{author}{\bibfnamefont{C.}~\bibnamefont{Kallin}},
  \bibinfo{journal}{Reports on Progress in Physics}
  \textbf{\bibinfo{volume}{75}}, \bibinfo{pages}{042501}
  (\bibinfo{year}{2012}).

\bibitem[{\citenamefont{Nandkishore}(2012)}]{2012_Nandkishore_SRO_PRB}
\bibinfo{author}{\bibfnamefont{R.}~\bibnamefont{Nandkishore}},
  \bibinfo{journal}{Phys. Rev. B} \textbf{\bibinfo{volume}{86}},
  \bibinfo{pages}{045101} (\bibinfo{year}{2012}).

\bibitem[{\citenamefont{Varma}(2000)}]{2000_Varma_HighTc_TRS_PRB}
\bibinfo{author}{\bibfnamefont{C.~M.} \bibnamefont{Varma}},
  \bibinfo{journal}{Phys. Rev. B} \textbf{\bibinfo{volume}{61}},
  \bibinfo{pages}{R3804} (\bibinfo{year}{2000}).

\bibitem[{\citenamefont{Chakravarty et~al.}(2001)\citenamefont{Chakravarty,
  Laughlin, Morr, and Nayak}}]{2001_Chakravarty_HighTc_TRS_PRB}
\bibinfo{author}{\bibfnamefont{S.}~\bibnamefont{Chakravarty}},
  \bibinfo{author}{\bibfnamefont{R.~B.} \bibnamefont{Laughlin}},
  \bibinfo{author}{\bibfnamefont{D.~K.} \bibnamefont{Morr}}, \bibnamefont{and}
  \bibinfo{author}{\bibfnamefont{C.}~\bibnamefont{Nayak}},
  \bibinfo{journal}{Phys. Rev. B} \textbf{\bibinfo{volume}{63}},
  \bibinfo{pages}{094503} (\bibinfo{year}{2001}).

\bibitem[{\citenamefont{Fauqu\'e et~al.}(2006)\citenamefont{Fauqu\'e, Sidis,
  Hinkov, Pailh\`es, Lin, Chaud, and Bourges}}]{2006_Fauque_HighTc_TRS_PRL}
\bibinfo{author}{\bibfnamefont{B.}~\bibnamefont{Fauqu\'e}},
  \bibinfo{author}{\bibfnamefont{Y.}~\bibnamefont{Sidis}},
  \bibinfo{author}{\bibfnamefont{V.}~\bibnamefont{Hinkov}},
  \bibinfo{author}{\bibfnamefont{S.}~\bibnamefont{Pailh\`es}},
  \bibinfo{author}{\bibfnamefont{C.~T.} \bibnamefont{Lin}},
  \bibinfo{author}{\bibfnamefont{X.}~\bibnamefont{Chaud}}, \bibnamefont{and}
  \bibinfo{author}{\bibfnamefont{P.}~\bibnamefont{Bourges}},
  \bibinfo{journal}{Phys. Rev. Lett.} \textbf{\bibinfo{volume}{96}},
  \bibinfo{pages}{197001} (\bibinfo{year}{2006}).

\bibitem[{\citenamefont{Folling et~al.}(2007)\citenamefont{Folling, Trotzky,
  Cheinet, Feld, Saers, Widera, Muller, and
  Bloch}}]{2007_Bloch_superlattice_Nature}
\bibinfo{author}{\bibfnamefont{S.}~\bibnamefont{Folling}},
  \bibinfo{author}{\bibfnamefont{S.}~\bibnamefont{Trotzky}},
  \bibinfo{author}{\bibfnamefont{P.}~\bibnamefont{Cheinet}},
  \bibinfo{author}{\bibfnamefont{M.}~\bibnamefont{Feld}},
  \bibinfo{author}{\bibfnamefont{R.}~\bibnamefont{Saers}},
  \bibinfo{author}{\bibfnamefont{A.}~\bibnamefont{Widera}},
  \bibinfo{author}{\bibfnamefont{T.}~\bibnamefont{Muller}}, \bibnamefont{and}
  \bibinfo{author}{\bibfnamefont{I.}~\bibnamefont{Bloch}},
  \bibinfo{journal}{Nature} \textbf{\bibinfo{volume}{448}},
  \bibinfo{pages}{1029} (\bibinfo{year}{2007}).

\bibitem[{\citenamefont{Tarruell et~al.}(2012)\citenamefont{Tarruell, Greif,
  Uehlinger, Jotzu, and Esslinger}}]{2011_Esslinger_OrbFermionDyn_Nature}
\bibinfo{author}{\bibfnamefont{L.}~\bibnamefont{Tarruell}},
  \bibinfo{author}{\bibfnamefont{D.}~\bibnamefont{Greif}},
  \bibinfo{author}{\bibfnamefont{T.}~\bibnamefont{Uehlinger}},
  \bibinfo{author}{\bibfnamefont{G.}~\bibnamefont{Jotzu}}, \bibnamefont{and}
  \bibinfo{author}{\bibfnamefont{T.}~\bibnamefont{Esslinger}},
  \bibinfo{journal}{Nature} \textbf{\bibinfo{volume}{483}},
  \bibinfo{pages}{302} (\bibinfo{year}{2012}).

\bibitem[{\citenamefont{\"Olschl\"ager
  et~al.}(2012)\citenamefont{\"Olschl\"ager, Wirth, Kock, and
  Hemmerich}}]{2012_Hemmerich_TopAvoidCross_PRL}
\bibinfo{author}{\bibfnamefont{M.}~\bibnamefont{\"Olschl\"ager}},
  \bibinfo{author}{\bibfnamefont{G.}~\bibnamefont{Wirth}},
  \bibinfo{author}{\bibfnamefont{T.}~\bibnamefont{Kock}}, \bibnamefont{and}
  \bibinfo{author}{\bibfnamefont{A.}~\bibnamefont{Hemmerich}},
  \bibinfo{journal}{Phys. Rev. Lett.} \textbf{\bibinfo{volume}{108}},
  \bibinfo{pages}{075302} (\bibinfo{year}{2012}).

\bibitem[{\citenamefont{Dhar et~al.}(2012)\citenamefont{Dhar, Maji, Mishra,
  Pai, Mukerjee, and Paramekanti}}]{2012_Dhar_ChiralMott_PRA}
\bibinfo{author}{\bibfnamefont{A.}~\bibnamefont{Dhar}},
  \bibinfo{author}{\bibfnamefont{M.}~\bibnamefont{Maji}},
  \bibinfo{author}{\bibfnamefont{T.}~\bibnamefont{Mishra}},
  \bibinfo{author}{\bibfnamefont{R.~V.} \bibnamefont{Pai}},
  \bibinfo{author}{\bibfnamefont{S.}~\bibnamefont{Mukerjee}}, \bibnamefont{and}
  \bibinfo{author}{\bibfnamefont{A.}~\bibnamefont{Paramekanti}},
  \bibinfo{journal}{Phys. Rev. A} \textbf{\bibinfo{volume}{85}},
  \bibinfo{pages}{041602} (\bibinfo{year}{2012}).

\bibitem[{\citenamefont{Dhar et~al.}(2013)\citenamefont{Dhar, Mishra, Maji,
  Pai, Mukerjee, and Paramekanti}}]{2013_Dhar_ChiralMott_PRB}
\bibinfo{author}{\bibfnamefont{A.}~\bibnamefont{Dhar}},
  \bibinfo{author}{\bibfnamefont{T.}~\bibnamefont{Mishra}},
  \bibinfo{author}{\bibfnamefont{M.}~\bibnamefont{Maji}},
  \bibinfo{author}{\bibfnamefont{R.~V.} \bibnamefont{Pai}},
  \bibinfo{author}{\bibfnamefont{S.}~\bibnamefont{Mukerjee}}, \bibnamefont{and}
  \bibinfo{author}{\bibfnamefont{A.}~\bibnamefont{Paramekanti}},
  \bibinfo{journal}{Phys. Rev. B} \textbf{\bibinfo{volume}{87}},
  \bibinfo{pages}{174501} (\bibinfo{year}{2013}).

\bibitem[{\citenamefont{{Zaletel} et~al.}(2013)\citenamefont{{Zaletel},
  {Parameswaran}, {R{\"u}egg}, and {Altman}}}]{2013_Altman_ChiralMott}
\bibinfo{author}{\bibfnamefont{M.~P.} \bibnamefont{{Zaletel}}},
  \bibinfo{author}{\bibfnamefont{S.~A.} \bibnamefont{{Parameswaran}}},
  \bibinfo{author}{\bibfnamefont{A.}~\bibnamefont{{R{\"u}egg}}},
  \bibnamefont{and} \bibinfo{author}{\bibfnamefont{E.}~\bibnamefont{{Altman}}},
  \bibinfo{journal}{ArXiv e-prints}  (\bibinfo{year}{2013}),
  \eprint{1308.3237}.

\bibitem[{\citenamefont{Binder}(1981)}]{1981_Binder}
\bibinfo{author}{\bibfnamefont{K.}~\bibnamefont{Binder}},
  \bibinfo{journal}{Zeitschrift f\"ur Physik B Condensed Matter}
  \textbf{\bibinfo{volume}{43}}, \bibinfo{pages}{119} (\bibinfo{year}{1981}).

\bibitem[{\citenamefont{Weber and Minnhagen}(1988)}]{1988_PRB_WeberMinnhagen}
\bibinfo{author}{\bibfnamefont{H.}~\bibnamefont{Weber}} \bibnamefont{and}
  \bibinfo{author}{\bibfnamefont{P.}~\bibnamefont{Minnhagen}},
  \bibinfo{journal}{Phys. Rev. B} \textbf{\bibinfo{volume}{37}},
  \bibinfo{pages}{5986} (\bibinfo{year}{1988}).

\bibitem[{\citenamefont{Killi and
  Paramekanti}(2012)}]{2012_Killi_CurrentQuench_PRA}
\bibinfo{author}{\bibfnamefont{M.}~\bibnamefont{Killi}} \bibnamefont{and}
  \bibinfo{author}{\bibfnamefont{A.}~\bibnamefont{Paramekanti}},
  \bibinfo{journal}{Phys. Rev. A} \textbf{\bibinfo{volume}{85}},
  \bibinfo{pages}{061606} (\bibinfo{year}{2012}).

\bibitem[{\citenamefont{Killi et~al.}(2012)\citenamefont{Killi, Trotzky, and
  Paramekanti}}]{2012_Killi_AnisotropicQuench_PRA}
\bibinfo{author}{\bibfnamefont{M.}~\bibnamefont{Killi}},
  \bibinfo{author}{\bibfnamefont{S.}~\bibnamefont{Trotzky}}, \bibnamefont{and}
  \bibinfo{author}{\bibfnamefont{A.}~\bibnamefont{Paramekanti}},
  \bibinfo{journal}{Phys. Rev. A} \textbf{\bibinfo{volume}{86}},
  \bibinfo{pages}{063632} (\bibinfo{year}{2012}).

\bibitem[{\citenamefont{Cai et~al.}(2012)\citenamefont{Cai, Duan, and
  Wu}}]{2011_Cai_UBEC}
\bibinfo{author}{\bibfnamefont{Z.}~\bibnamefont{Cai}},
  \bibinfo{author}{\bibfnamefont{L.-M.} \bibnamefont{Duan}}, \bibnamefont{and}
  \bibinfo{author}{\bibfnamefont{C.}~\bibnamefont{Wu}}, \bibinfo{journal}{Phys.
  Rev. A} \textbf{\bibinfo{volume}{86}}, \bibinfo{pages}{051601}
  (\bibinfo{year}{2012}).

\bibitem[{\citenamefont{Seibold and Lorenzana}(2001)}]{2001_Seibold_TGutz_PRL}
\bibinfo{author}{\bibfnamefont{G.}~\bibnamefont{Seibold}} \bibnamefont{and}
  \bibinfo{author}{\bibfnamefont{J.}~\bibnamefont{Lorenzana}},
  \bibinfo{journal}{Phys. Rev. Lett.} \textbf{\bibinfo{volume}{86}},
  \bibinfo{pages}{2605} (\bibinfo{year}{2001}).

\bibitem[{\citenamefont{Snoek and Hofstetter}(2007)}]{2007_Snoek_TGutzwiller}
\bibinfo{author}{\bibfnamefont{M.}~\bibnamefont{Snoek}} \bibnamefont{and}
  \bibinfo{author}{\bibfnamefont{W.}~\bibnamefont{Hofstetter}},
  \bibinfo{journal}{Phys. Rev. A} \textbf{\bibinfo{volume}{76}},
  \bibinfo{pages}{051603} (\bibinfo{year}{2007}).

\bibitem[{\citenamefont{Anderlini et~al.}(2007)\citenamefont{Anderlini, Lee,
  Brown, Sebby-Strabley, Phillips, and Porto}}]{2007_NIST_superlattice_Nature}
\bibinfo{author}{\bibfnamefont{M.}~\bibnamefont{Anderlini}},
  \bibinfo{author}{\bibfnamefont{P.~J.} \bibnamefont{Lee}},
  \bibinfo{author}{\bibfnamefont{B.~L.} \bibnamefont{Brown}},
  \bibinfo{author}{\bibfnamefont{J.}~\bibnamefont{Sebby-Strabley}},
  \bibinfo{author}{\bibfnamefont{W.~D.} \bibnamefont{Phillips}},
  \bibnamefont{and} \bibinfo{author}{\bibfnamefont{J.~V.} \bibnamefont{Porto}},
  \bibinfo{journal}{Nature} \textbf{\bibinfo{volume}{448}},
  \bibinfo{pages}{452} (\bibinfo{year}{2007}).

\end{thebibliography}
\bibliographystyle{apsrev}

\end{document}